\documentclass[journal,draftcls,onecolumn,12pt,twoside]{IEEEtran}
\IEEEoverridecommandlockouts
\usepackage{amsmath}
\usepackage{algorithmic}
\usepackage{amssymb}
\usepackage[ruled]{algorithm2e}
\usepackage{booktabs}
\usepackage{color}
\usepackage{cite}
\usepackage{cuted}
\usepackage{enumitem}
\usepackage{graphicx}
\usepackage{float}
\usepackage{flushend}
\usepackage{lipsum} 
\usepackage{mathrsfs} 
\usepackage{multirow}
\usepackage{stfloats}
\usepackage{subcaption}

\newtheorem{theorem}{\textbf{Theorem}}
\newtheorem{corollary}{Corollary}
\newtheorem{assumption}{\textbf{Assumption}}
\newtheorem{lemma}{\textbf{Lemma}}

\newtheorem{proposition}{\textbf{Proposition}}

\newtheorem{definition}{\textbf{Definition}}

\def\BibTeX{{\rm B\kern-.05em{\sc i\kern-.025em b}\kern-.08em
    T\kern-.1667em\lower.7ex\hbox{E}\kern-.125emX}}
\begin{document}

\title{An Index Policy for Minimizing the Uncertainty-of-Information of Markov Sources
}

\author{Gongpu~Chen,
	Soung Chang Liew,~\IEEEmembership{Fellow,~IEEE}	
	\thanks{This work was supported in part by the General Research Funds (Project No. 14200221) established under the University Grant Committee of the Hong Kong Special Administrative Region, China. (\textit{Corresponding author: Soung-Chang Liew})}
	\thanks{The authors are with the Department of Information Engineering, The Chinese University of Hong Kong, Shatin, Hong Kong (E-mail: \{gpchen, soung\}@ie.cuhk.edu.hk).}
	
}

\maketitle
\vspace{-10mm}

\begin{abstract}
This paper focuses on the information freshness of finite-state Markov sources, using the uncertainty of information (UoI) as the performance metric. Measured by Shannon’s entropy, UoI can capture not only the transition dynamics of the Markov source but also the different evolutions of information quality caused by the different values of the last observation.
We consider an information update system with $M$  finite-state Markov sources transmitting information to a remote monitor via $m$  communication channels ($1\le m <M$). At each time, only $m$  Markov sources can be selected to transmit their latest information to the remote monitor. Our goal is to explore the optimal scheduling policy to minimize the sum-UoI of the Markov sources. The problem is formulated as a restless multi-armed bandit (RMAB). We relax the RMAB and then decouple the relaxed problem into $M$  single bandit problems. Importantly, analyzing the single bandit problem provides useful properties with which the relaxed problem reduces to maximizing a concave and piecewise linear function, allowing us to develop a gradient method to solve the relaxed problem and obtain its optimal policy. By rounding up the optimal policy for the relaxed problem, we obtain an index policy for the original RMAB problem. 
Notably, the proposed index policy is universal in the sense that it applies to general RMABs with bounded cost functions. Moreover, we show that our policy is asymptotically optimal as $m$ and $M$ tend to $\infty$ with $m/M$ fixed.
In non-asymptotic cases, numerical results demonstrate that our index policy is near-optimal and performs as well as the celebrated Whittle index policy in the problems that are Whittle-indexable. Unlike the Whittle index policy, our index policy does not require ``indexability"; the indices can be computed regardless of indexability in the Whittle's sense. Thus, our index policy is a promising alternative method for the class of RMABs of concern: it can be used when the Whittle index policy is not viable and it performs as well as the Whittle index policy even when the Whittle index policy is viable. 

\end{abstract}


\section{Introduction}
The demand for real-time information delivery has increased sharply with the growing deployment of modern monitoring and control systems over the past years. Developing communication techniques to support such real-time information delivery is a key issue for the next-generation communication systems. A recent research topic is to use the age of information (AoI), instead of classical latency, as the performance metric for information update systems. AoI is a metric of information freshness proposed in 2011-2012 \cite{AoI_2011,AoI_originalpaper}. It measures the time elapsed since the generation of the last packet received at the receiver. A growing body of work has adopted AoI as the performance metric for communication in real-time applications \cite{AoIdesign_Kadota2019,AoI2016Costa,AoICoding_2022TII,AoIdesign_henry2020,sombabu2022whittle,AoI_Cao2023}, resulting in different system designs than using the classical latency metric.

\begin{figure}[!t]
	\centering
	\begin{subfigure}[b]{0.4\linewidth}
		\includegraphics[width=\linewidth]{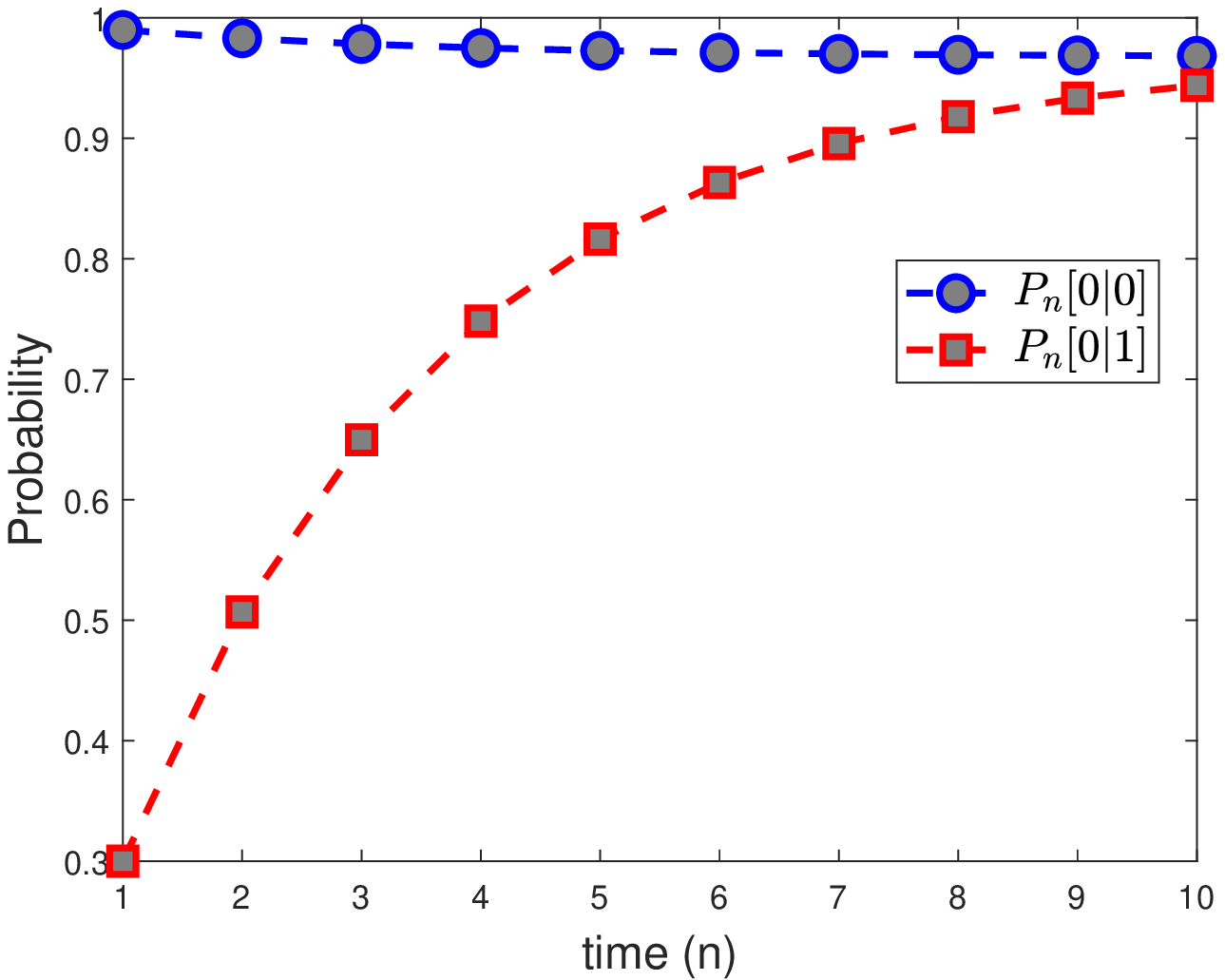}
		\caption{$n$-step transition probability.}
		\label{fig:bi_belief}
	\end{subfigure}
	\begin{subfigure}[b]{0.4\linewidth}
		\includegraphics[width=\linewidth]{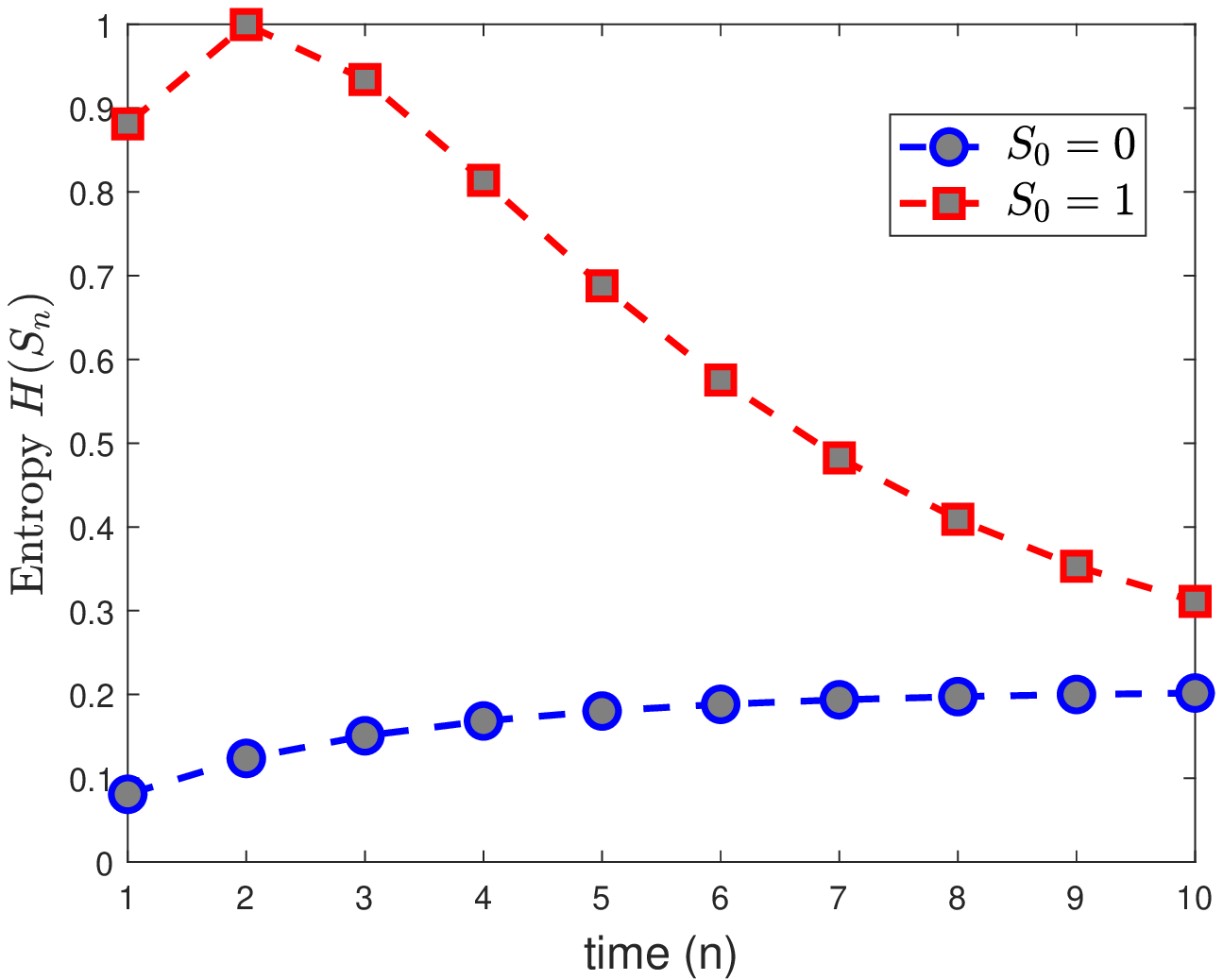}
		\caption{Entropy over time.}
		\label{fig:bi_UoI}
	\end{subfigure}
	\caption{Information quality evolves with time at different rates, given different values of the last observation. (a): $P_n[0|0]$ and $P_n[0|1]$ denote the $n$-step transition probabilities to state 0 from states 0 and 1, respectively. (b): Evolution of UoI of the Markov chain with different values of the last observation $S_0$.}
	\label{fig:example}
	\vspace{-10mm}
\end{figure}

Besides the original AoI metric, many studies also considered various cost functions of AoI to capture the system dynamics \cite{AoIcost_nonlinear_MIT,AoIcost_nonlinear2019,Aoicost_nonlinear,AoIaware18}. All these AoI-related metrics assume that the evolution of information quality only depends on the “age” of information. However, in many practical cases, different contents of the last updated information may result in the quality of information evolving with time at different rates. In particular, some information (even from the same source) may age more quickly than other information. For example, consider a binary Markov chain, say with states 0 and 1, being observed by a remote monitor. Assume that the transition probabilities of the Markov chain are $P[1|0]=0.01$ and $P[0|1]=0.3$, and they are known to the monitor. Upon receiving the state $S_0$ of the Markov chain at time $0$, the information quality of $S_0$ at the monitor may evolve with time in a way that depends on the value of $S_0$. As shown in Fig. \ref{fig:bi_belief}, if $S_0=0$, then the Markov chain remains in state $0$ with a high probability in the time that follows; hence $S_0=0$ is still useful in the next few time steps. In contrast, if $S_0=1$, then the monitor can hardly infer the states of the process in the next few time steps without new observations; hence $S_0=1$ quickly becomes outdated. In general, AoI-related metrics fail to reflect the different rates of information quality evolution caused by different values of the last observation.

A better way to look at the above example is to interpret information quality from the perspective of uncertainty measured by Shannon’s entropy. Such motivated, the concept of uncertainty of information (UoI) was proposed in \cite{UoI_Gongpu} as a metric of information freshness for binary Markov chains.
In particular, let $S_n$ denote the state of the binary Markov chain at time step $n$ and assume that the latest state update of the chain in the monitor is $S_0$, then the UoI of the Markov chain at time $n>0$ is the entropy of $S_n$ conditioned on the value of $S_0$. Note that $S_n$ is unknown (hence a random variable) to the monitor at time $n$  given that $S_0$ is the latest state update. UoI measures how much we do not know about the current state of the Markov chain in the lack of new observations, with the value of the last observation taken into account. This metric provides a better understanding of the above example, as shown in Fig. \ref{fig:bi_UoI}. The information quality of $S_0=0$ decays slowly over time because the UoI conditioned on  $S_0=0$ remains low for the next few time steps. In contrast, the observation $S_0=1$  quickly becomes useless because it leads to large UoIs in the next few time steps. This example also suggests an interesting observation that information quality does not always evolve with time in a monotonically increasing way. This is another motivation for studying the UoI metric. From the perspective of AoI, it seems strange at first glance that information quality may increase with time. However, as shown in the example, given a specific observation, the uncertainty of the real-time state in the monitor may decrease with time.

This paper adopts UoI as the metric of information freshness and studies the minimum-UoI scheduling problem. In particular, we consider an information update system in which $M$  finite-state Markov chains transmit their instantaneous states to a remote monitor via $m$  communication channels ($1\le m <M$ ). We assume that each channel can be used by only one Markov chain at each time step. Scheduling refers to selecting $m$  chains at each time to update their states. We formulate the scheduling problem as a restless multi-armed bandit (RMAB) \cite{gittins_RMAB_book} and study the optimal scheduling policy to minimize the sum-UoI of the $M$  Markov chains.

\subsection{Literature Review}
Many studies in the literature investigated information freshness from the information-theoretic perspective \cite{MutualInform_sunyin,VoI_HiddenMC,VoI_entropy,Leiyin2020,Coding_MarkovSource,Sampling_Markov}. For example, the work in \cite{MutualInform_sunyin} studied the optimal sampling policy for a binary Markov source and used the mutual information between the real-time state and the delivered samples as the metric of information freshness. It turns out that the mutual information is a non-decreasing function of AoI and is independent of the value of the last sample; because mutual information treats the last sample as a random variable and does not consider its realization. A similar idea was explored in \cite{VoI_HiddenMC} for hidden Markov models, where the information freshness is evaluated by the mutual information between the current state and the sequence of observed measurements. The authors in \cite{VoI_entropy} proposed to use the differential entropy computed by the Fisher information matrix as a measure of information for sampling continuous-time stochastic processes and investigated the optimal sampling policy.
A noteworthy AoI-related metric is the age of incorrect information (AoII) proposed in \cite{AoII_maatouk2020}. It partially considers the information content by proposing age penalties only when the new update differs from the receiver's prediction; such updates are called informative. However, AoII still fails to reflect the different rates of information quality evolution caused by the particular content of informative updates. 

The most relevant work to the current paper is \cite{UoI_Gongpu} authored by us, where we proposed the concept of UoI for binary Markov chains and studied the optimal scheduling policy to minimize the average UoI. As pointed out in \cite{UoI_Gongpu}, if the information source is a symmetric binary Markov chain, then UoI reduces to a non-decreasing cost function of AoI. However, in the general asymmetric case, UoI evolves with time in a way that is affected by the values of received observations, while AoI does not. Most of the existing studies on information freshness only investigated binary Markov sources, while the results about the general finite-state Markov sources are still very lacking. 
This paper generalizes the work in \cite{UoI_Gongpu} in three ways. First, we extend the definition of UoI from binary Markov chains to general finite-state Markov chains and study the minimum-UoI scheduling problem. Second, \cite{UoI_Gongpu} assumes that the communication channels are reliable, while this paper generalizes the setting to include the case of unreliable channels. These two extensions bring a key challenge, as will be elaborated on later. Finally, this paper investigates the optimal policies for both expected total discounted UoI and long-term average UoI, while \cite{UoI_Gongpu} only studied the latter objective. 

 RMAB is a widely used stochastic scheduling model, which fits our UoI-scheduling problem perfectly. It is a sequential decision-making problem in which a set of resources must be allocated to $m$ out of $M$ ($1\le m <M$) processes at each time step in a way that maximizes the expected gain over the whole time horizon. Conventionally, each process is referred to as a bandit \cite{gittins_RMAB_book}. Since RMAB is PSPACE-hard \cite{RMAB_PSPACEhard}, the optimal policy is generally unavailable. 
The Whittle index policy \cite{RMAB_Whittle1988} is probably the most well-known method to tackle RMABs heuristically due to its simple form and asymptotic optimality \cite{weber1990index}. It computes an index for each state of every bandit and then selects the $m$ bandits with the largest $m$ indices at each time step (hence, the term ``index policy'').  However, the Whittle index policy applies only to the RMABs that possess the “indexability” property. Although some studies investigated sufficient conditions for the indexability of RMABs \cite{nino_2001,nino2007dynamic,Whittle_app2006}, those conditions are strict and can be satisfied by only a limited class of RMABs. In general, establishing the indexability of an RMAB is challenging and demands an arduous analytical effort \cite{Shiling2020,Liu2010,Whittle_app2018,villar_2016}. Exacerbating the situation is that, even if an RMAB is indexable, computing the Whittle index may still be complex \cite{Index_compute2020}. 

In fact, \cite{UoI_Gongpu} established the indexability for the UoI scheduling with binary Markov sources and reliable channels, and then developed an iterative algorithm to compute the Whittle index. The results, however, are not applicable to the general case of finite-state Markov sources studied in this paper. Proving that the active set of the optimal policy for the single bandit problem shrinks monotonically with the service charge is a core step in establishing indexability. In the binary case, UoI is a function on the 1-dimensional probability simplex. The optimal policy for the single bandit problem has a threshold structure, which is essential in establishing indexability. For the Markov sources with $N$  states, UoI is a function on the  $(N-1)$-dimensional probability simplex, so is the value function of the single bandit problem. Consequently, the active set of the optimal policy for the single bandit problem is a subset of the  $(N-1)$-dimensional probability simplex. It is tremendously challenging to analyze how the multi-dimensional active set varies with the service charge; hence the indexability can not be proved, and the Whittle index policy cannot be used as a result. 
{In addition, even for the UoI scheduling of binary Markov sources, whether the indexability holds for the general setting of unreliable channels is still an open problem.}
Such motivated, we propose for the UoI scheduling problem a novel index policy that does not require indexability.

\subsection{Main Results}
In this paper, we investigate the minimum-UoI scheduling of $M$  finite-state Markov sources. Two commonly used objectives are considered, i.e., the expected total discounted UoI and the long-term average UoI. We formulate the problem as an RMAB and develop a novel index policy that does not require indexability in the traditional sense. In particular, our index policy computes an index for each state of a bandit (our RMAB has $M$  bandits, each associated with a Markov source). The index of each bandit reflects the gain of selecting this bandit to transmit at the current time slot.
The policy selects the $m$  bandits with the largest $m$  indices at each time slot. We thus call it gain index policy. Our development consists of four steps. 

The first step is relaxation and decomposition. Specifically, we relax the problem and then convert it into an “unconstrained problem” using the Lagrange multiplier method and linear programming (LP) theory. Doing so yields a sup-min problem without constraints. The basic idea is to solve the resulting sup-min problem and then derive an index policy for the original RMAB by rounding up the optimal policy for the sup-min problem. Toward this end, we fix the decision variable, denoted by $\lambda$, of the sup problem and decouple the min problem into multiple subproblems called single bandit problems. Each single bandit problem is a belief Markov decision process (MDP) associated with a Markov source.

The second step is analyzing the single bandit problem to yield useful properties. A key result is that the value function of the single bandit problem is an increasing, concave, and piecewise linear function of  $\lambda$. On this basis, the sup-min problem reduces to maximizing a concave and piecewise linear function of  $\lambda$. 

The third step is solving the sup-min problem. We propose a gradient method to compute the optimal $\lambda$  and prove that the method can stop in finite steps and return a  $\lambda$ satisfying $|\lambda  - {\lambda ^*}| < \epsilon$  for any $\epsilon>0$, where $\lambda^*$ denotes the optimal $\lambda$.

The fourth step is determining the index policy for the original RMAB and computing the indices. The optimal policy for the sup-min problem (hence the relaxed problem) can be determined with $\lambda^*$. Rounding up the optimal policy leads to an index policy for the original RMAB. We put forth an efficient algorithm to compute the indices of each bandit.

{Surprisingly, the gain index policy is asymptotically optimal as $m$ and $M$ tend to $\infty$ with $m/M$ fixed. We establish the asymptotic optimality by showing that the gain index policy behaves almost the same as the optimal policy for the relaxed problem in the limiting case. }
The results in this paper apply to a large class of RMABs, with the UoI-scheduling problem (the particular focus of this paper) as a special case. In fact, a general RMAB with bounded cost function admits our method. 

The rest of this paper is organized as follows. Section \ref{sec:System} presents the system model and the RMAB formulation of the UoI scheduling problem. Section \ref{sec:dec&single} decouples the RMAB with the discounted UoI criterion and studies the resulting single bandit problem. Section \ref{sec:policy} derives the gain index policy for the RMAB. Section \ref{sec:average} extends the results to the UoI scheduling with the average UoI criterion. Section \ref{sec:asy-opt} establishes the asymptotic optimality of the policy for both criteria. Section \ref{sec:sim} presents simulation results that demonstrate the excellent performance of the proposed index policy. Finally, Section \ref{sec:con} concludes this paper.

\subsection{Notations}
For a positive integer $M$, let $[M]\triangleq \{1,\cdots,M \}$ denote the set of integers between $1$ and $M$. $P[\cdot | \cdot]$ denotes the conditional probability. $\mathbb{R}^N$ and $\mathbb{R}^{N\times N}$ denote the $N$-dimensional vector space and the set of $N\times N$ matrices, respectively. The set of positive integers is denoted by $\mathbb{N}=\{1,2,3,\cdots\}$. $\mathbf{I}$ denotes the identity matrix with appropriate dimensions. For any discrete set $\mathcal{S}$ and a function defined on this set, say $V:\mathcal{S}\to \mathbb{R}$, we use $V=[V(s)]_{s\in \mathcal{S}}$ to denote vector form of this function.

\section{Problem Statement} \label{sec:System}
\subsection{System Model}
Consider a system in which a central monitor observes $M$  remote discrete-time Markov chains (DTMC). The states of the remote DTMCs can be delivered to the central monitor via $m$  wireless channels ($1\le m <M$). These channels operate in a time-slotted manner, and each time slot corresponds to a time step of the remote DTMCs. A channel can be used for only one DTMC at each time slot to avoid interference. We thus need a scheduling policy that selects $m$  DTMCs at each time slot to update their states. At the beginning of each time slot, a selected DTMC sends its instantaneous state to the central monitor; the monitor receives the state information by the end of the slot if the transmission succeeds. The channels are potentially unreliable due to wireless channel fading. Assume that each transmission from the  $i$-th DTMC to the central monitor succeeds with probability ${\rho _i} \in (0,1],i \in [M]$, regardless of the channel used.

For simplicity, we assume that each of the DTMCs has $N$ states\footnote{The results in this paper apply to the general case that the DTMCs have different state dimensions.}, where $2\le N <\infty$. Let ${S_i}(t) \in [N],i \in [M]$, denote the state of the $i$-th DTMC at time $t$. For each $i\in [M]$, $S_i(t)$ evolves in time according to a transition matrix ${{\bf{T}}^{(i)}} \in {\mathbb{R}^{N \times N}}$. That is,
\begin{align*}
P[{S_i}(t + 1) = k|{S_i}(t) = n] = {\bf{T}}_{nk}^{(i)},\quad n,k \in [N],
\end{align*}
where ${\bf{T}}_{nk}^{(i)}$ denotes the $(n,k)$-th entry of matrix ${{\bf{T}}^{(i)}}$. We also assume that all the DTMCs are aperiodic and irreducible. 

UoI for binary Markov chains is defined in \cite{UoI_Gongpu} as a metric of information freshness. In this paper, we extend the definition of UoI to general finite-state Markov chains. In particular, denote by $U_i(t)$  the UoI of the  $i$-th DTMC at the end of time slot $t$. Then $U_i(t)$ is the entropy of $S_i(t+1)$ conditioned on the latest observation of the $i$-th DTMC at the central monitor. Note that the end of time slot $t$  is equivalent to the beginning of time slot $t+1$  in continuous time. Formally, let $\Omega$ denote the $(N-1)$-dimensional probability simplex, i.e.,
\begin{align*}
\Omega  \buildrel \Delta \over = \left\{ {x \in {\mathbb{R}^N}:\sum\limits_{i = 1}^N {{x_i}}  = 1,{x_i} \ge 0{\text{ for all }}i} \right\}.
\end{align*}
Then the Shannon entropy of a distribution $x\in \Omega$  is given by
\begin{align*}
H(x) =  - \sum\limits_{i = 1}^N {{x_i}{{\log }_2}{x_i}}. 
\end{align*}
Suppose that the latest observation of the  $i$-th DTMC at time $t$ is $S_i(t')$, where $t'\le t$. Then $U_i(t)$ is given by
\begin{align*}
{U_i}(t) =  - \sum\limits_{s \in [N]} {P[{S_i}(t + 1) = s|{S_i}(t')]{{\log }_2}} P[{S_i}(t + 1) = s|{S_i}(t')].
\end{align*}
Intuitively, $U_i(t)$  measures how much the central monitor does not know about the latest state of the $i$-th DTMC.
Note that $U_i(t)$  not only depends on the age of the latest observation (i.e., $t+1-t'$), but also depends on the value of the latest observation $S_i(t')$, making UoI a different metric from the cost functions of AoI.

Since only $m$  remote DTMCs can deliver their current states to the central monitor at each time slot, there is a trade-off among the UoIs of the remote DTMCs. In this paper, we are interested in the scheduling of the updates of the remote DTMCs.  Two commonly used objectives are considered. The first one is to minimize the expected total discounted UoI of the $M$  DTMCs:
\begin{align} \label{eq:disObj}
\min\quad E\left[ {\sum\limits_{t = 1}^\infty  {\sum\limits_{i = 1}^M {{\beta ^{t - 1}}{U_i}(t)} } |\chi } \right],
\end{align}
where $\beta \in [0,1)$ is the discount factor, $\chi  = [{\chi ^{(1)}}, \cdots ,{\chi ^{(M)}}]$ and ${\chi ^{(i)}} \in {\mathbb{R}^N}$ is the distribution of $S_i(1)$. We will refer to \eqref{eq:disObj} as UoI-scheduling with the discounted cost criterion. { Discounting arises naturally in applications in which we account for the time value of UoIs. It also fits the case where the monitor aims to minimize the expected total UoI over a time horizon whose length is random and independent of the monitor \cite{puterman1994markov}. }

The other objective is to minimize the expected average UoI over the infinite horizon:
\begin{align}
\min \quad {\rm{  }}E\left[ {\mathop {\lim }\limits_{n \to \infty } \frac{1}{n}\sum\limits_{t = 1}^n {\sum\limits_{i = 1}^M {{U_i}(t)} } |\chi } \right].
\end{align}
The above is referred to as UoI-scheduling with the average cost criterion. This objective is suitable for the case where UoIs are of equal importance over the whole time horizon.

\subsection{RMAB Formulation}
We formulate a belief MDP for each DTMC, based on which UoI is a concave function of the belief state. Then the scheduling problem can be formulated as an RMAB. Specifically, each belief MDP is a bandit of the RMAB. At every time step, we need to allocate the communication channels to $m$ out of $M$ bandits in a way that minimizes the sum UoI of all bandits. The formal definition of this RMAB problem is presented below. 

The belief MDP associated with the $i$-th DTMC is defined by the belief state space $\mathcal{S}^{(i)}\subseteq \Omega$, the action space $\{0,1\}$, the transition kernel of belief states, the cost function (i.e., UoI) $H(\cdot)$, and the discount factor $\beta$.
Denote by $u_i(t)\in \{0,1\}$ the action applied to the $i$-th DTMC at time $t$, where $u_i(t)=1$ means that the $i$-th DTMC is selected to update in slot $t$ and $u_i(t)=0$ otherwise. We also refer to $u_i(t)=1$  as the active action and $u_i(t)=0$  as the passive action. Further, let  $\gamma_i(t)\in \{0,1\}$  indicate whether the central monitor receives an observation of the  $i$-th DTMC at the end of time slot  $t$. In particular, $\gamma_i(t)=1$  if and only if $u_i(t)=1$  and the transmission succeeds.

The belief state $X_i(t)\in\mathcal{S}^{(i)}$  for the  $i$-th DTMC is defined as the distribution of $S_i(t)$  at the beginning of time slot $t$  conditioned on the latest observation $S_i(t')$, where $t' < t$. That is, the $k$-th element of $X_i(t)$, denoted by $x_{i,k}(t)$, is the following conditional probability: 
\begin{align*}
{x_{i,k}}(t) = P\left[ {{S_i}(t) = k|{S_i}(t')} \right],\quad i \in [M],k \in [N].
\end{align*}
For each $i\in [M]$, the process $\{ {X_i}(t):t = 1,2,3 \cdots \} $ is Markovian. Specifically, the belief state $X_i(t)$ evolves according to $\gamma_i(t)$  and $S_i(t)$  in the following way:
\begin{align} \label{eq:XtTrans}
{X_i}(t + 1) =  \begin{cases}
{\bf{T}}_k^{(i)}, &{\text{if }}{\gamma _i}(t) = 1{\text{ and }}{S_i}(t) = k,\\
{{\bf{T}}^{(i)}}{X_i}(t),&{\text{if }}{\gamma _i}(t) = 0.
\end{cases} 
\end{align}
where ${\bf{T}}_k^{(i)}$ is the $k$-th column of matrix ${\bf{T}}^{(i)}$. Note that we have assumed ${X_i}(1) = {\chi ^{(i)}}$. Hence the set of all possible values of $X_i(t)$ (i.e., the belief state space $\mathcal{S}^{(i)}$) is a proper subset of $\Omega$. We defer the rigorous definition of the belief state space to the next section.

Based on the definition of belief state, the UoI of the  $i$-th DTMC at the end of time slot $t$  can be written as ${U_i}(t) = H\left( {{X_i}(t + 1)} \right)$. Now, the UoI-scheduling with the discounted cost criterion is formulated as an RMAB and can be formally written as
\begin{align} \label{eq:P1-obj}
{\text{P1: }}\min \quad &E\left[ {\sum\limits_{t = 1}^\infty  {\sum\limits_{i = 1}^M {{\beta ^{t - 1}}H\left( {{X_i}(t)} \right)} } |\chi } \right]\\ \label{eq:P1-st1}
{\rm{        }}s.t.\quad &\sum\limits_{i = 1}^M {{u_i}(t) = m,\quad \forall t} \\  \label{eq:P1-st2}
&{u_i}(t) \in \{ 0,1\} ,\quad \forall i,t.
\end{align}
Likewise, the UoI-scheduling with the average cost criterion can be formulated as
\begin{align} \label{eq:P2}
{\text{P2: }}\min \quad &E\left[ {\mathop {\lim }\limits_{n \to \infty } \frac{1}{n}\sum\limits_{t = 1}^n {\sum\limits_{i = 1}^M {H\left( {{X_i}(t)} \right)} } |\chi } \right]\\ \notag
{\rm{        }}s.t. \quad & \eqref{eq:P1-st1},\eqref{eq:P1-st2}.
\end{align}
We will first study the UoI-scheduling with the discounted cost criterion and develop an index policy for this problem. Then we extend the results to the case of the average cost criterion and use a similar method to establish an index policy for problem P2.

\section{Decomposition and The Single Bandit Problem} \label{sec:dec&single}
In this section, we consider a relaxation of problem P1. To solve this relaxed problem, we apply the Lagrange multiplier method and then make a decomposition, leading to $M$ single bandit problems. We then study the single bandit problems and establish properties that are useful in the subsequent development. Throughout this section, the discount factor $\beta$  is fixed and $\beta\in [0,1)$.

\subsection{Relaxation and Decomposition}
Constraint \eqref{eq:P1-st1} means that exactly $m$  DTMCs can be selected to transmit in each time slot. A general method proposed by Whittle \cite{RMAB_Whittle1988} is to relax the constraint to the following:
\begin{align} \label{eq:RelSt}
E\left[ {\sum\limits_{t = 1}^\infty  {{\beta ^{t - 1}}} \mathop \sum \limits_{i = 1}^M {u_i}(t)} \right] = \sum\limits_{t = 1}^\infty  {{\beta ^{t - 1}}} m = \frac{m}{{1 - \beta }}.
\end{align}
Clearly, \eqref{eq:RelSt} is a looser constraint on $\{u_i(t) \}$ than \eqref{eq:P1-st1}. Hence replacing constraint \eqref{eq:P1-st1} by \eqref{eq:RelSt} leads to a relaxation of problem P1. Using the Lagrange multiplier method \cite{bertsekas1997nonlinear}, we can convert the relaxed problem into an unconstrained problem by introducing a Lagrange multiplier $\lambda$:
\begin{align} \label{eq:P1a}
{\text{P1(a): }}\mathop {\min }\limits_{\{ {u_i}(t)\} } \mathop {\sup }\limits_{\lambda  \ge 0} {\rm{  E}}\left[ {\sum\limits_{t = 1}^\infty  {\sum\limits_{i = 1}^M {{\beta ^{t - 1}}\big[ {H\left( {{X_i}(t)} \right) + \lambda {u_i}(t)} \big]} |\chi } } \right] - \frac{{m\lambda }}{{1 - \beta }}.
\end{align}
For any fixed $\lambda$, the term $E[ \cdot |\chi ]$  in problem P1(a) can be considered as the expected total discounted cost of an MDP with cost function $H(X)+\lambda u$ (cf. \eqref{eq:P1-obj}). The Lagrange multiplier $\lambda$  can be interpreted as a service charge for applying the active action to a bandit. It is well-known that an MDP can be formulated as an LP. We thus can transform the relaxed problem into an LP. Since $H(\cdot)$  is a bounded function, the relaxed problem must have a finite value function. By Slater’s condition, an LP has strong duality if the problem is strictly feasible and has a bounded optimal value \cite{boyd2004convex}. This implies that the min and sup in \eqref{eq:P1a} can be interchanged. That is, problem P1(a) is equivalent to the following:
\begin{align} \label{eq:P1b}
{\text{P1(b): }}\mathop {\sup }\limits_{\lambda  \ge 0} \mathop {\min }\limits_{\{ {u_i}(t)\} } {\rm{  E}}\left[ {\sum\limits_{t = 1}^\infty  {\sum\limits_{i = 1}^M {{\beta ^{t - 1}}\left[ {H\left( {{X_i}(t)} \right) + \lambda {u_i}(t)} \right]} |\chi } } \right] - \frac{{m\lambda }}{{1 - \beta }}.
\end{align}
To solve the above problem, we first fix $\lambda$  and study the inner minimization problem. As we will prove later, the optimal value of the minimization problem is a concave function of $\lambda$. We thus can further convert the problem to a concave program.

For a fixed $\lambda$, problem P1(b) can be decoupled into $M$  subproblems:
\begin{align} \label{eq:Ji}
{J_i}(\lambda ): = \mathop {\min }\limits_{\{ {u_i}(t)\} } {\rm{  E}}\left[ {\sum\limits_{t = 1}^\infty  {{\beta ^{t - 1}}\left[ {H\left( {{X_i}(t)} \right) + \lambda {u_i}(t)} \right]|{\chi ^{(i)}}} } \right],\quad i \in [M].
\end{align}
We refer to ${J_i}(\lambda )$  as a single bandit problem with service charge $\lambda$. In particular, ${J_i}(\lambda )$  is to minimize the expected total discounted cost of a belief MDP associated with the  $i$-th DTMC. The belief MDP is similar to that defined in Section II.B; the only difference is that the cost incurred by state-action pair $(X,u)$ is $H(X)+\lambda u$  instead of $H(X)$. We thus interpret $\lambda$  as a service charge of taking active action; it is incurred every time the active action is taken.

\subsection{The Single Bandit Problem} \label{subsec:disc-single}
In this part, we focus on the single bandit problem defined in \eqref{eq:Ji}. Since all the  single bandit problems are of the same form, we will consider an arbitrary bandit and, for simplicity, drop the bandit index from all notations in this part. For example, ${J_i}(\lambda )$  and ${{\bf{T}}^{(i)}}$  will be written as ${J}(\lambda )$  and  ${{\bf{T}}}$, respectively. 

The single bandit problem is a belief MDP with belief state space ${\cal S} \buildrel \Delta \over = \left\{ {{\bf{T}}_k^n:k \in [N],n \in \mathbb{N} } \right\}$ and action space $\{0,1\}$. We remind the reader that ${\bf{T}}_k^n$  is the  $k$-th column of the  $n$-step transition matrix of the remote DTMC; it should not be confused with ${\bf{T}}_k^{(i)}$. From \eqref{eq:XtTrans}, the belief state transition probability from  $X(t) = [{x_1} \cdots {x_N}]$ to $X(t+1)$  is given by
\begin{align} \label{eq:SbTrans}
P[X(t + 1)|X(t),u(t)] =  \begin{cases}
\rho {x_k},&{\text{  if }}u(t) = 1,X(t + 1) = {{\bf{T}}_k}\\
1 - \rho ,&{\text{ if }}u(t) = 1,X(t + 1) = {\bf{T}}X(t)\\
1,&{\text{       if }}u(t) = 0,X(t + 1) = {\bf{T}}X(t)\\
0,&{\text{       otherwise.}}
\end{cases} 
\end{align}
Recall that $\rho$ is the transmission success probability. We will use  $J(\lambda)$ to denote this belief MDP with service charge $\lambda$. The optimal policy for the single bandit problem can be determined by the Bellman equation, for $X \in {\cal S}$:
\begin{align} \label{eq:Bellman}
V(X,\lambda ) = H(X) + \min \left\{ {\lambda  + \beta \rho \mathop \sum \limits_{i = 1}^N {x_i}V({\mathbf{T}_i},\lambda ) + \beta (1 - \rho )V(\mathbf{T} X,\lambda ),\beta V(\mathbf{T}X,\lambda )} \right\},
\end{align}
where $x_i$  is the  $i$-th element of  $X$. We express the value function as $V(X,\lambda)$  to emphasize its dependency on $\lambda$. In the context that  $\lambda$ is fixed, we may also simply write the value function as  $V(X)$. Note that the belief state space is a countable set and  ${\cal S} \subset \Omega $. We find it is convenient to extend the belief state space from ${\cal S}$  to  $\Omega $. That is, we consider the Bellman equation \eqref{eq:Bellman} for $X\in\Omega$. By doing this, $V(X)$  can be treated as a function defined on a continuous space. All the results established based on this treatment are also valid over ${\cal S}$.

In the following, we will establish some useful properties of the optimal policy and the value function. They are essential for us to develop an index policy for the original RMAB problem.

\begin{lemma} \label{Lem:Vxconcave}
	For any fixed $\lambda \ge 0$, the value function $V(X)$ is a concave function of $X\in \Omega$.
\end{lemma}
\begin{IEEEproof}
	This result can be proved via value iteration. See Appendix I for the complete proof.
\end{IEEEproof}

The concavity of $V(X)$  allows us to establish a structural property of the optimal policy. Define the active function  $a(X,\lambda)$ and passive function $r(X,\lambda)$  as
\begin{align} \label{eq:actFun}
a(X,\lambda ) &\buildrel \Delta \over = \lambda  + \beta \rho \mathop \sum \limits_{i = 1}^N {x_i}V({{\bf{T}}_i},\lambda ) + \beta (1 - \rho )V({\bf{T}}X,\lambda ),\\ \label{eq:pasFun}
r(X,\lambda ) &\buildrel \Delta \over = \beta V({\bf{T}}X,\lambda ).
\end{align}
Similarly, we may write the above functions as $a(X)$  and $r(X)$  when  $\lambda$ is fixed. Then the Bellman equation can be written as
\begin{align}
V(X) = H(X) + \min \{ a(X),r(X)\}. 
\end{align}
Clearly, it is optimal to take active action in a belief state  $X$ if  $r(X) \ge a(X)$. Specially, if  $r(X) = a(X)$, then it is equally optimal to take the two actions in $X$. We next show that the optimal policy has a special structure.
\begin{lemma} \label{Lem:ConvexAct}
	For any fixed $\lambda\in [0,\infty)$, the optimal policy for the single bandit problem is a convex-sampling policy. In particular, there exists a convex set $\mathcal{A}_\lambda$, such that $r(X) \ge a(X)$ for $X\in \mathcal{A}_\lambda$ and $r(X) < a(X)$ for $X\notin \mathcal{A}_\lambda$. The optimal policy is given by
	\begin{align*}
	u(t) =  \begin{cases}
	1,&{\text{ if }X}(t) \in {\mathcal{A}_\lambda }\\
	0,&{\text{ otherwise.}}
	\end{cases} 
	\end{align*}
	where  $\mathcal{A}_\lambda$ is referred to as the active set. 
\end{lemma}
\begin{IEEEproof}
This lemma is proved based on Lemma 1. See Appendix I for the details.
\end{IEEEproof}

The above two lemmas examine the optimal policy and the value function of the single bandit problem with a particular service charge $\lambda$. In the following, let us turn our focus to how the value of  $\lambda$ affects the value function and the optimal policy. 

\begin{theorem} \label{thm:Vlambda}
	For any $X\in \Omega$, $V(X,\lambda)$ is an increasing, concave, and piecewise linear function of $\lambda \in [0,\infty)$. In addition,
	\begin{align*}
	0 \le \frac{{\partial V(X,\lambda )}}{{\partial \lambda }} \le \frac{1}{{1 - \beta }}.
	\end{align*}
\end{theorem}
\begin{IEEEproof}
	Throughout this proof, we consider $\lambda \in [0,\infty)$.
	We first show that $V(X,\lambda)$ is increasing w.r.t. $\lambda$. By definition, $V(X,\lambda)$  is the expected total discounted cost incurred by the optimal policy, given that the initial state is $X$. That is, $V(X,\lambda)$  can be expressed as
	\begin{align} \notag
	V(X,\lambda ) &= E\left[ {\sum\limits_{t = 1}^\infty  {{\beta ^{t - 1}}\left[ {H\left( {X(t)} \right) + \lambda u(t)} \right]|X} } \right]\\ \label{eq:lem3-1}
	&= E\left[ {\sum\limits_{t = 1}^\infty  {{\beta ^{t - 1}}H\left( {X(t)} \right)|X} } \right] + \lambda E\left[ {\sum\limits_{t = 1}^\infty  {{\beta ^{t - 1}}u(t)|X} } \right],
	\end{align}
	where $E[\cdot|X]$ is the expectation taken over the Markov chain generated by the optimal policy with initial state $X$. It follows that
	\begin{align} \label{eq:lem3-2}
	\frac{{\partial V(X,\lambda )}}{{\partial \lambda }} = E\left[ {\sum\limits_{t = 1}^\infty  {{\beta ^{t - 1}}u(t)|X} } \right] \le \sum\limits_{t = 1}^\infty  {{\beta ^{t - 1}} = \frac{1}{{1 - \beta }}} .
	\end{align}
	The inequality is satisfied with equality if the optimal policy takes active action in all belief states, i.e., $u(t)=1$  for all $t$. Likewise, since $u(t)\in\{0,1\}$, we also have
	\begin{align} \label{eq:lem3-3}
	\frac{{\partial V(X,\lambda )}}{{\partial \lambda }} = E\left[ {\sum\limits_{t = 1}^\infty  {{\beta ^{t - 1}}u(t)|X} } \right] \ge 0.
	\end{align}
	The inequality is satisfied with equality if the optimal policy takes passive action in all belief states, i.e., $u(t)=0$  for all $t$. The above inequality implies that $V(X,\lambda)$ is an increasing function of $\lambda$.
	
	We next show that $V(X,\lambda)$ is concave w.r.t. $\lambda$. Assume $0 \le \lambda_1 < \lambda_2$. For an arbitrary $\theta \in [0,1]$, let $\lambda' = \theta {\lambda _1} + (1 - \theta ){\lambda _2}$. Furthermore, suppose that policy  $\pi$ is the optimal policy for the single bandit problem with service charge $\lambda'$, i.e., $J(\lambda')$. If policy $\pi$ is applied to the bandit with service charge ${\lambda _1} = \lambda ' + (1 - \theta )({\lambda _1} - {\lambda _2})$, then the expected total discounted cost is given by
	\begin{align} \notag
	{V_\pi }(X,{\lambda _1}) &\buildrel \Delta \over = {E_\pi }\left[ {\sum\limits_{t = 1}^\infty  {{\beta ^{t - 1}}\left[ {H\left( {X(t)} \right) + (\lambda ' + (1 - \theta )({\lambda _1} - {\lambda _2}))u(t)} \right]|X} } \right]\\ \label{eq:lem3-4}
	&= V(X,\lambda ') + (1 - \theta )({\lambda _1} - {\lambda _2}){E_\pi }\left[ {\sum\limits_{t = 1}^\infty  {{\beta ^{t - 1}}u(t)|X} } \right].
	\end{align}
	where $E_\pi [\cdot]$ means that the expectation is taken over the Markov chain induced by policy $\pi$. Since $V(X,\lambda_1)$ denotes the optimal value function of the single bandit problem $J(\lambda_1)$, we have
	\begin{align} 
	{V_\pi }(X,{\lambda _1}) \ge V(X,{\lambda _1}).
	\end{align}
	Likewise, applying policy $\pi$ to $J(\lambda_2)$ yields
	\begin{align} \label{eq:lem3-5}
	{V_\pi }(X,{\lambda _2}) = V(X,\lambda ') - \theta ({\lambda _1} - {\lambda _2}){E_\pi }\left[ {\sum\limits_{t = 1}^\infty  {{\beta ^{t - 1}}u(t)|X} } \right] \ge V(X,{\lambda _2}).
	\end{align}
	From \eqref{eq:lem3-4}-\eqref{eq:lem3-5}, we can verify that
	\begin{align}
	\theta V(X,{\lambda _1}) + (1 - \theta )V(X,{\lambda _2}) \le V(X,\theta {\lambda _1} + (1 - \theta ){\lambda _2}).
	\end{align}
	Therefore, $V(X,\lambda)$ is a concave function of $\lambda$.
	
	Finally, we prove the piecewise linearity of $V(X,\lambda)$  by showing that its partial derivative is piecewise-constant. According to \eqref{eq:lem3-2}, the partial derivative is piecewise constant if, for any $\lambda$, there exists an interval $I_\lambda$  containing  $\lambda$ such that the optimal policy is invariant in this interval. We show this by contradiction.  Denote by $\Pi$ the set of optimal policies for the single bandit problem with service charge $\lambda$ (for some $\lambda$ and $X$ such that $a(X,\lambda)=r(X,\lambda)$, it is equally optimal to take passive action and active action in $X$; hence there may be multiple optimal policies).
	Then, let  $\Pi_\sigma$ denote the optimal policy for the single bandit problem $J(\lambda + \sigma)$. Suppose that there exists a $\lambda$ such that $\Pi_\sigma$ and $\Pi$ are disjoint for $|\sigma|$ arbitrarily small. Then for any ${\pi } \in {\Pi  }$ and ${\pi _\sigma } \in {\Pi _\sigma }$,
	\begin{align} \notag
	V(X,\lambda  + \sigma ) - V(X,\lambda )
	&=\sigma {E_{{\pi _\sigma }}}\left[ {\sum\limits_{t = 1}^\infty  {{\beta ^{t - 1}}u(t)|X} } \right] + {E_{{\pi _\sigma }}}\left[ {\sum\limits_{t = 1}^\infty  {{\beta ^{t - 1}}\left[ {H\left( {X(t)} \right) + \lambda u(t)} \right]|X} } \right] \\ 
	&- {E_\pi }\left[ {\sum\limits_{t = 1}^\infty  {{\beta ^{t - 1}}\left[ {H\left( {X(t)} \right) + \lambda u(t)} \right]|X} } \right].
	\end{align}
	By assumption, $\Pi_\sigma$ and $\Pi$ are disjoint, hence ${\pi _\sigma } \notin {\Pi }$ and there exists a nonempty set of  $X$ such that
	\begin{align}
	{E_{{\pi _\sigma }}}\left[ {\sum\limits_{t = 1}^\infty  {{\beta ^{t - 1}}\left[ {H\left( {X(t)} \right) + \lambda u(t)} \right]|X} } \right] - {E_\pi }\left[ {\sum\limits_{t = 1}^\infty  {{\beta ^{t - 1}}\left[ {H\left( {X(t)} \right) + \lambda u(t)} \right]|X} } \right] \ne 0.
	\end{align}
	The above implies that $\max_X|V(X,\lambda  + \sigma ) - V(X,\lambda )|$ does not go to zero as $\sigma \to 0$. But this contradicts the fact that $V(X,\lambda)$  is continuous w.r.t. $\lambda$  for any $X\in \Omega$. Therefore, the single bandit problems $J(\lambda)$  and $J(\lambda + \sigma)$  must share at least one identical optimal policy when $|\sigma|$  is small enough; then $\partial V/\partial \lambda $  is a constant in a nonempty interval containing $\lambda$. We thus conclude that $V(X,\lambda)$  is a piecewise linear function of $\lambda$. 
	
\end{IEEEproof}

Theorem 1 states that the partial derivative of the value function w.r.t. $\lambda$  is bounded. We also discussed in the proof that the upper (lower) bound is achieved if there exists a  $\lambda$ such that the optimal policy is to take active (passive) action in all belief states. The corollary below states that the upper and lower bounds are achievable.

\begin{corollary} \label{Coro:lambda0}
	For the optimal policy for the single bandit problem $J(\lambda)$,
	\begin{itemize}
		\item [1.] If $\lambda = 0$, the optimal policy is to take active action in all belief states.
		\item [2.] There exists a $\bar{\lambda} > 0$ such that for any $\lambda > \bar{\lambda}$, the optimal policy is to take passive action in all belief states. 
	\end{itemize} 
\end{corollary}
\begin{IEEEproof}
	Statement 1 follows easily from Lemma 2. Statement 2 can be proved based on an argument used in the proof of Theorem 1. See Appendix I for the complete proof.
\end{IEEEproof}

Corollary 1 implies an interesting result about the UoI metric. As shown by the example of Fig. \ref{fig:example}, UoI may decrease with AoI. One thus may suspect that it is better for a Markov source not to transmit new observations when its UoI is decreasing with time. However, statement 1 of the above corollary means that the active action is always better than the passive action when the service charge $\lambda=0$. Therefore, a Markov source should send new observations at every available opportunity to minimize its expected total discounted UoI. 

In the above discussions, we extended the belief state space from the countable set $\mathcal{S}$ to the probability simplex $\Omega$  to facilitate analysis. Let us now return to the original belief state space ${\cal S} =  \left\{ {{\bf{T}}_k^n:k \in [N],n \in \mathbb{N} } \right\}$ and consider some issues related to practical computations. Since  $\mathcal{S}$ contains infinite states, solving the single bandit problem by classical methods (e.g., value iteration and policy iteration) is computationally intractable. Fortunately, the convergence feature of the belief states allows us to construct a finite-state MDP to approximate the original one. In particular, denote by $\omega\in \mathbb{R}^N$ the equilibrium distribution of the remote DTMC, i.e.,
\begin{align} \label{eq:steadyDistr}
\mathop {\lim }\limits_{n \to \infty } {\bf{T}}_k^n = \omega ,\quad \forall k \in [N].
\end{align}
Note that $\omega \in \mathcal{S}$. We also call $\omega$ the equilibrium belief state. We next construct a finite-state MDP to approximate the single bandit problem $J(\lambda)$.

\begin{definition}[Truncated belief MDP]
	For a given positive integer $L$, the $L$-truncated belief MDP of $J(\lambda)$, denoted by $J^L(\lambda)$, is defined as follows:
	\begin{itemize}
		\item State space: ${{\cal S}^L} \buildrel \Delta \over = \left\{ {\omega ,{\bf{T}}_k^n:k \in [N],n \in [L]} \right\}$. 
		\item Action space: $\{0,1\}$.
		\item Cost of state-action pair $(X,u) \in {{\cal S}^L} \times \{ 0,1\} $: $H(X)+\lambda u$.
		\item Discount factor: $\beta$.
		\item Transition rule: the probability of transitioning from $X(t) \in {{\cal S}^L}$ to $X(t+1) \in {{\cal S}^L}$ with action $u(t)$ is the same as that of $J(\lambda)$ given in \eqref{eq:SbTrans}, except that
		\begin{align*}
		P[X(t + 1) = \omega |X(t) = {\bf{T}}_k^L,u(t) = 0] = 1,\quad k \in [N].
		\end{align*}
	\end{itemize}
\end{definition}

Given a large integer $L$, belief states $\{ {\bf{T}}_k^{L + i}:k \in [N],i \in \mathbb{N} \} $ are already close to $\omega$, hence the truncated belief MDP just treats them as being $\omega$ to approximate $J(\lambda)$.
We will use ${\phi ^L}(X,\lambda )$  to denote the optimal value function of $J^L(\lambda)$. Clearly, $J^L(\lambda)$  has  $NL+1$ states. Hence it is computationally tractable by classical methods when $N$  and  $L$ are not too large. The theorem below provides a bound for the approximation error of $J^L(\lambda)$. 

\begin{theorem} \label{thm:errorbound1}
	Let $L$ be a positive integer such that
	\begin{align*}
	\mathop {\max }\limits_{i \in [N]} ||{\bf{T}}_i^L - \omega |{|_\infty } \le {\eta _L}{\text{ and }}\mathop {\max }\limits_{i \in [N],k \ge 0} \left| {H({\bf{T}}_i^{L + k}) - H(\omega )} \right| \le {\sigma _L},
	\end{align*}
	where ${\eta _L},{\sigma _L} > 0$, $|| \cdot |{|_\infty }$ denotes the max norm. In addition, suppose that $H(X)\le B_H$ for $X\in \mathcal{S}$. Then for any $\lambda\in [0,\infty)$ and $X\in \mathcal{S}^L$,
	\begin{align*}
	|V(X,\lambda ) - {\phi ^L}(X,\lambda )| \le \frac{{\beta {\sigma _L}}}{{1 - \beta }} + \beta \rho {\eta _L}N\frac{{{B_H} + \lambda }}{{{{\left( {1 - \beta } \right)}^2}}}.
	\end{align*}
\end{theorem}
\begin{IEEEproof}
	We present a sketch of the proof here and defer the full proof to Appendix II. We first construct an auxiliary MDP, denoted by $J'(\lambda)$. The state space and action space of  $J'(\lambda)$ are the same as that of $J(\lambda)$; the transition rule and cost function for $X\in \mathcal{S}^L$  are also identical to that of $J(\lambda)$. For $X\in \mathcal{S} - \mathcal{S}^L$, $J'(\lambda)$ has different transition rules and cost functions from $J(\lambda)$: for any state-action pair $(X,u) \in ({\cal S} - {\cal S}{^L}) \times \{ 0,1\} $, the cost is $H(\omega)+\lambda u$ and
	\begin{align*}
	P[X(t + 1)|X(t)=X,u(t)=u] =  \begin{cases}
	\rho {\omega _k},&{\text{  if }}u = 1,X(t + 1) = {{\bf{T}}_k}\\
	1 - \rho ,&{\text{ if }}u = 1,X(t + 1) = {\bf{T}}X\\
	1,&{\text{       if }}u = 0,X(t + 1) = {\bf{T}}X\\
	0,&{\text{       otherwise.}}
	\end{cases}
	\end{align*}
	That is, all $X \in {\cal S} - {\cal S}{^L}$ have the same cost function and transition rule as the equilibrium belief state $\omega$ (Note that $\mathbf{T}\omega = \omega$). In effect, each belief state in $ {\cal S} - {\cal S}{^L}$ is equivalent to $\omega$. Hence the set $ {\cal S} - {\cal S}{^L}$ can be aggregated with state $\omega$, and this does not change the values of other states. In other words, let $\varphi (X)$ denote the value function of $J'(\lambda)$, then $\varphi (X)=\phi^L(X)$ for $X \in {{\cal S}^L}$.
	
	Since $J'(\lambda)$ and $J(\lambda)$ have the same state space and action space, a policy of $J(\lambda)$  can also be applied to $J'(\lambda)$. For a policy $\pi$ of $J(\lambda)$, denote by $V_\pi$ and $\varphi_\pi$ the value functions of $J(\lambda)$ and $J'(\lambda)$ under this policy, respectively.
	To establish the desired result, we first bound  $|V_\pi - \varphi_\pi|$ for arbitrary $\pi$. Then it can be proved that $|V - \varphi|$ admits the same bound. Finally, using the fact that  $\varphi (X)=\phi^L(X)$ for $X \in {{\cal S}^L}$ yields the desired bound for $|V-\phi^L|$.
\end{IEEEproof}

According to \eqref{eq:steadyDistr}, $\eta_L$ can be arbitrarily small as long as $L$  is large enough. In this case, the continuity of $H(X)$  implies that $\sigma_L$  can also be arbitrarily small. Consequently, the upper bound in Theorem \ref{thm:errorbound1} tends to zero as $L\to \infty$; hence $J^L(\lambda)$ could well approximate $J(\lambda)$ when $L$ is large. Since $\lambda$  appears in the second term of the bound, one may concern about the tightness of the bound in the case that $\lambda$ is pretty large. According to Corollary 1, when $\lambda$  is large enough, the optimal policy for  $J(\lambda)$ is to take passive action in all belief states. A similar argument could show that this is also true for $J^L(\lambda)$. Under the policy of taking passive action in all states, $\lambda$ has no effect on the value function. Namely, for $\lambda$ large enough, the bound in Theorem \ref{thm:errorbound1} reduces to a form that is independent of $\lambda$: 
\begin{align}
|V(X,\lambda ) - {\phi ^L}(X,\lambda )| \le \frac{{\beta {\sigma _L}}}{{1 - \beta }} + \beta \rho {\eta _L}N\frac{{{B_H}}}{{{{\left( {1 - \beta } \right)}^2}}}.
\end{align}
Theorem \ref{thm:errorbound1} allows us to approximate the single bandit problem $J(\lambda)$  by a finite-state MDP $J^L(\lambda)$. This property is useful in the practical computation of our index policy for the RMAB.

\section{The Gain Index Policy}  \label{sec:policy}
In this section, we return to the relaxed problem P1(b). We first develop an iterative algorithm to find the optimal solution for P1(b) based on the properties of the single bandit problem established in the previous section. By rounding up the optimal solution for the relaxed problem, we obtain an index policy for the original RMAB problem.

\subsection{The Optimal Policy for the Relaxed Problem} \label{subsec:Dpolicy-A}
For $i\in [M]$ and a fixed $\lambda\ge 0$, denote by $V_i(X,\lambda)$ the optimal value function of the single bandit problem associated with the $i$-th DTMC (i.e., $J_i(\lambda)$). That is,
\begin{align}
{V_i}({\chi ^{(i)}},\lambda ) = \mathop {\min }\limits_{\{ {u_i}(t)\} } {\rm{  E}}\left[ {\sum\limits_{t = 1}^\infty  {{\beta ^{t - 1}}\left[ {H\left( {{X_i}(t)} \right) + \lambda {u_i}(t)} \right]|{\chi ^{(i)}}} } \right].
\end{align}
Define a function for $\lambda  \in [0,\infty )$:
\begin{align}
f(\lambda ) \buildrel \Delta \over = \sum\limits_{i = 1}^M {{V_i}({\chi ^{(i)}},\lambda )}  - \frac{{m\lambda }}{{1 - \beta }}.
\end{align}
It follows immediately from Theorem \ref{thm:Vlambda} that $f(\lambda)$  is a piecewise linear and concave function of $\lambda$. Then the relaxed problem P1(b) can be converted to a concave program as follows:
\begin{align}
{\text{P1(c): }}\mathop {\sup }\limits_{\lambda  \ge 0} \left\{ {\sum\limits_{i = 1}^M {{V_i}({\chi ^{(i)}},\lambda )}  - \frac{{m\lambda }}{{1 - \beta }}} \right\} = \mathop {\sup }\limits_{\lambda  \ge 0} f(\lambda ).
\end{align}

We can use a gradient method to solve problem P1(c). In particular, let $\lambda_0 = 0$  and define the following iteration:
\begin{align} \label{eq:gradient}
{\lambda _{k + 1}} = {\lambda _k} + {a_k}f'({\lambda _k}),\quad k = 0,1,2, \cdots .
\end{align}
where $\{a_k\}$  is the sequence of stepsize,  $f'(\cdot)$ is the derivative of $f(\cdot)$. 
The piecewise linearity and concavity of $f(\lambda)$  imply two facts: (1) problem P1(c) must have optimal solutions; if the optimal solution is not unique, then there exists an interval, say $[{\lambda ^l},{\lambda ^u}]$, within which all $f(\lambda)$ are optimal.
(2)  $f(\lambda)$ is not differentiable everywhere. In the points that $f'(\lambda)$  is not defined, we can use the right (or left) derivative, which always exists. The details will be elaborated on later.
It turns out that, by choosing a proper sequence of stepsize, the sequence  $\{\lambda_k\}$ generated by \eqref{eq:gradient} converges to an optimal solution of problem P1(c). The lemma below implies a stopping criterion for the above gradient method.

\begin{lemma} \label{Lem:f'}
	Denote by $\lambda^*$ an optimal solution to P1(c), then $\lambda^* > 0$ and
	\begin{align*}
	f'(\lambda _ - ^*) \ge 0,\quad f'(\lambda _ + ^*) \le 0.
	\end{align*}
\end{lemma}
\begin{IEEEproof}
	The derivative of $f(\lambda)$ is
	\begin{align} \label{eq:f'}
	\frac{{df(\lambda )}}{{d\lambda }} = \sum\limits_{i = 1}^M {\frac{{\partial {V_i}({\chi ^{(i)}},\lambda )}}{{\partial \lambda }}}  - \frac{m}{{1 - \beta }} = \sum\limits_{i = 1}^M {E\left[ {\sum\limits_{t = 1}^\infty  {{\beta ^{t - 1}}{u_i}(t)|{\chi ^{(i)}}} } \right]}  - \frac{m}{{1 - \beta }}.
	\end{align}
	According to Corollary 1, if $\lambda=0$, the optimal policy is to take active action in all belief states. It follows that the summation term in \eqref{eq:f'} is equal to $M/(1-\beta)$ when $\lambda = 0$. On the other hand, as $\lambda \to \infty$, the optimal policy takes passive action in all belief states. Hence the summation term in \eqref{eq:f'} tends to $0$ as  $\lambda \to \infty$. We thus have
	\begin{align} \label{eq:df0}
	\frac{{df(0)}}{{d\lambda }} = \frac{M}{{1 - \beta }} - \frac{m}{{1 - \beta }} > 0,\quad \mathop {\lim }\limits_{\lambda  \to \infty } \frac{{df(\lambda )}}{{d\lambda }} =  - \frac{m}{{1 - \beta }} < 0.
	\end{align}
	According to Theorem \ref{thm:Vlambda}, ${V_i}({\chi ^{(i)}},\lambda )$ is increasing, concave, and piecewise linear w.r.t. $\lambda$. Hence $df/d\lambda$  is piecewise-constant and decreasing w.r.t. $\lambda$. Then \eqref{eq:df0} implies that there must exist a  $\lambda^* >0$ such that
	\begin{align} \label{eq:f'lambda}
	\frac{{df(\lambda _ - ^*)}}{{d\lambda }} \ge 0,\quad \frac{{df(\lambda _ + ^*)}}{{d\lambda }} \le 0.
	\end{align}
	The above inequalities are satisfied with equality if there exists an interval $(\lambda^l,\lambda^u)$ such that $f'(\lambda)=0$ for $\lambda\in (\lambda^l,\lambda^u)$. In this case, any  $\lambda^*\in [\lambda^l,\lambda^u]$  is an optimal solution. Function $f(\lambda)$  achieves its maximum value in $\lambda^*$. The desired result follows immediately from \eqref{eq:f'} and \eqref{eq:f'lambda}.
		
\end{IEEEproof}

\begin{figure}[!t]
	\centering
	\includegraphics[width=3.1in]{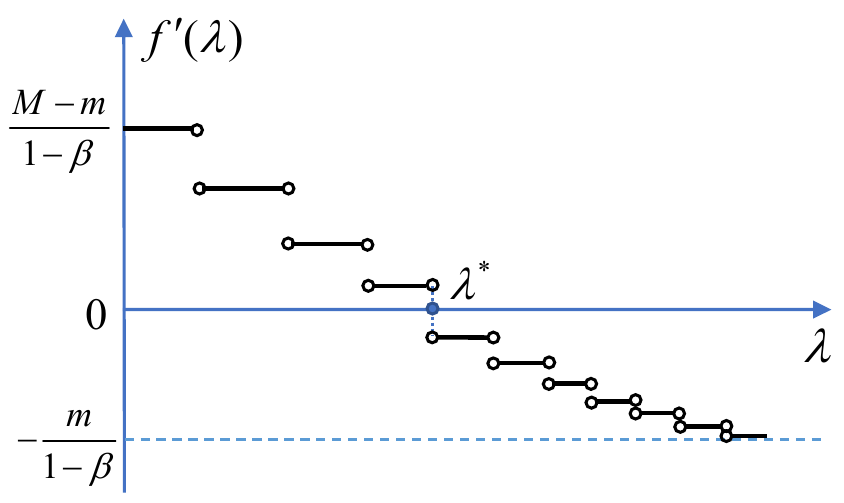}
	\caption{Derivative of $f(\lambda)$ and the optimal solution $\lambda^*$.}
	\label{fig:df}
	\vspace{-10mm}
\end{figure}
Fig. \ref{fig:df} is a sketch of $f'(\lambda)$. As mentioned in the above proof, it is also possible that $f'(\lambda)=0$ in an interval $(\lambda^l,\lambda^u)$, in which case any $\lambda^*\in [\lambda^l,\lambda^u]$  is an optimal solution, and the inequalities in Lemma \ref{Lem:f'} satisfy the equality. According to Lemma \ref{Lem:f'}, we can define a stopping criterion for the gradient method defined in \eqref{eq:gradient} as follows:
\begin{align} \label{eq:stopCrit}
f'({\lambda_k})f'({\lambda _{k + 1}}) \le 0{\text{ and }}|{\lambda _{k + 1}} - {\lambda _k}| < \epsilon,
\end{align}
where $\epsilon>0$. The above condition is satisfied only if $\lambda^*$ is in between $\lambda_k$ and $\lambda_{k+1}$. The theorem below shows that, given any $\epsilon>0$, the above condition can be satisfied within a finite number of iterations. When the iteration stops, we obtain a  $\lambda_{k}$ such that $|{\lambda ^*} - {\lambda _k}| < \epsilon$.

\begin{theorem} \label{Thm:finitestop}
	Choosing $a_k = c/k$ with $c$ being a positive constant, and let $\{\lambda_{k}\}$ be the sequence generated by \eqref{eq:gradient}. Then given any $\epsilon>0$, there exists a finite integer $D_\epsilon$ such that
	\begin{align*}
	f'({\lambda _{{D_\epsilon}}})f'({\lambda _{{D_\epsilon} + 1}}) \le 0{\text{ and }}|{\lambda _{{D_\epsilon} + 1}} - {\lambda _{{D_\epsilon}}}| < \epsilon.
	\end{align*}
	The interval between ${\lambda _{{D_\epsilon} + 1}}$ and ${\lambda _{{D_\epsilon} }}$ contains at least one optimal solution of P1(c).
	
\end{theorem}
\begin{IEEEproof}
	Given $\lambda_0=0$ and $a_k = c/k$, define
	\begin{align}
	D \triangleq \min \left\{ {k:|{\lambda _{k + 1}} - {\lambda _k}| < \epsilon} \right\}.
	\end{align}
	Note that $f'(\lambda)$ is bounded, and $\{a_k\}$ is a decreasing sequence that converges to 0. Hence $D$ must be finite. Further, $f(\lambda)$ is piecewise linear indicates that $f'(\lambda)$  is discontinuous and piecewise constant. As discussed just below Fig. \ref{fig:df}, there are two possibilities: (1) $f'(\lambda)\neq 0$ for all $\lambda >0$, and $\lambda^*$ is the unique point satisfying $f'(\lambda_-^*) > 0$ and $f'(\lambda_+^*) < 0$; (2) $f'(\lambda)=0$ in some interval $(\lambda^l,\lambda^u)$, then $ [\lambda^l,\lambda^u]$ is the set of optimal solutions. We will prove the theorem for both cases.
	
	(1) if $f'(\lambda)\neq 0$ for all $\lambda >0$, then we must have $\lambda_{D+1} \neq \lambda_D$. First, assume that  $\lambda_{D+1} > \lambda_D$. Then we have three subcases:
	\begin{itemize}
		\item [i).] $\lambda_{D+1} > \lambda^* > \lambda_D$. It is clear that $f'({\lambda _{D + 1}}) > 0$ and $f'({\lambda _{D}}) < 0$. Hence $D_\epsilon = D$.
		\item[ii).] $\lambda_{D+1} > \lambda_D > \lambda^*$. Then $f'({\lambda _{D + 1}}) \le f'({\lambda _D}) < 0$, implying that ${\lambda _{D + 1}} = {\lambda _D} + {a_D}f'({\lambda _D}) < {\lambda _D}$. We obtain a contradiction. Hence this situation is impossible.
		\item[iii).] $\lambda^* > \lambda_{D+1} > \lambda_D$. Then $f'({\lambda _D}) \ge f'({\lambda _{D + 1}}) \ge f'(\lambda _ - ^*) > 0$. Let
		\begin{align*}
		q \buildrel \Delta \over = \max \left\{ {k:k > D,{\lambda _k} < {\lambda ^*}} \right\}.
		\end{align*}
		Then $f'({\lambda _D}) \ge f'({\lambda _k}) \ge f'(\lambda _ - ^*)$ for all $D\le k \le q$. We use contradiction to show that $q$ is finite.  Suppose that $q$ is infinite, then it means that $\lambda_k < \lambda^*$ for all $k\ge D$. Note that for $n\ge D$,
		\begin{align}
		{\lambda _{n + 1}} = {\lambda _D} + \sum\limits_{k = D}^n {{a_k}f'({\lambda _k})}  = {\lambda _D} + c\sum\limits_{k = D}^n {\frac{1}{k}f'({\lambda _k})}  \ge {\lambda _D} + cf'(\lambda _ - ^*)\sum\limits_{k = D}^n {\frac{1}{k}}.
		\end{align}
		It is well-known that $\sum\nolimits_{k = D}^n {1/k} $ diverges as $n\to \infty$. Then $cf'(\lambda _ - ^*) > 0$ implies that there must exist a finite integer $n$  such that
		\begin{align}
		cf'(\lambda _ - ^*)\sum\limits_{k = D}^n {\frac{1}{k}}  > {\lambda ^*} - {\lambda _D}.
		\end{align}
		Consequently, ${\lambda _{n + 1}} > {\lambda ^*}$. We obtain a contradiction; hence $d$  must be finite. Now, we have ${\lambda _{q + 1}} > {\lambda ^*} > {\lambda _q}$, which implies that
		\begin{align} \label{eq:thm3-39}
		f'({\lambda _q}) \ge f'(\lambda _ - ^*) > 0,f'({\lambda _{q + 1}}) \le f'(\lambda _ + ^*) < 0.
		\end{align}
		Further, $0<a_q<a_D$ and $0 < f'({\lambda _q}) \le f'({\lambda _D})$ imply that
		\begin{align} \label{eq:thm3-40}
		{\lambda _{q + 1}} - {\lambda _q} = {a_q}f'({\lambda _q}) < {a_D}f'({\lambda _D}) = {\lambda _{D + 1}} - {\lambda _D} < \epsilon.
		\end{align}
		From \eqref{eq:thm3-39} and \eqref{eq:thm3-40}, it is clear that $D_\epsilon = q <\infty$. 		
	\end{itemize}
Applying a similar argument can prove that the desired result also holds in the case of $\lambda_{D + 1} < \lambda_D$.

(2) if $f'(\lambda)=0$ in some interval $(\lambda^l,\lambda^u)$. It is possible that ${\lambda _{D + 1}} \in ({\lambda ^l},{\lambda ^u})$. In this case, we have $f'(\lambda_{D + 1})=0$ and $D_\epsilon = D$. We next consider the nontrivial case that  ${\lambda _{D + 1}} \notin ({\lambda ^l},{\lambda ^u})$. It follows that $\lambda_{D + 1}\neq \lambda_{D}$. To see this, note that $\lambda_{D + 1} = \lambda_D$ if and only if $f'(\lambda_D)=0$, while this is possible only if ${\lambda _{D + 1}} = \lambda_D \in ({\lambda ^l},{\lambda ^u})$.

As in the previous case, we only examine the case of $\lambda_{D + 1} > \lambda_D$, and the proof for the case of $\lambda_D > \lambda_{D + 1} $ will be omitted for simplicity. Consider the following subcases:
\begin{itemize}
	\item [i).] ${\lambda _{D + 1}} > {\lambda ^u} > {\lambda ^l} > {\lambda _D}$. It is clear that $f'({\lambda _{D + 1}}) > 0$ and $f'({\lambda _{D}}) < 0$. Hence $D_\epsilon = D$.
	\item[ii).] ${\lambda _{D + 1}} > {\lambda _D} > {\lambda ^u}$. Then $f'({\lambda _{D + 1}}) \le f'({\lambda _D}) < 0$, implying ${\lambda _{D + 1}} = {\lambda _D} + {a_D}f'({\lambda _D}) < {\lambda _D}$. We obtain a contradiction. Hence this situation is impossible.
	\item[iii).] ${\lambda ^l} > {\lambda _{D + 1}} > {\lambda _D}$. Then $f'({\lambda _D}) \ge f'({\lambda _{D + 1}}) \ge f'(\lambda_- ^l) > 0$. We modify the definition of $q$ accordingly:
	\begin{align*}
	q \buildrel \Delta \over = \max \left\{ {k:k > D,{\lambda _k} < {\lambda ^l}} \right\}.
	\end{align*}
	Applying a similar argument as before, we can show that $q$  is finite and
	\begin{align*}
	f'({\lambda _q})f'({\lambda _{q + 1}}) \le 0{\text{ and }}{\lambda _{q + 1}} - {\lambda _q} < \epsilon.
	\end{align*}
	Therefore, $D_\epsilon = q<\infty$ and $\lambda^l \in (\lambda_{{D_\epsilon}},\lambda_{{D_\epsilon} + 1})$. This completes the proof. 
\end{itemize}
	
\end{IEEEproof}

From the above discussions, we can compute the optimal solution to problem P1(c) by the gradient method as long as $f'(\lambda)$  can be evaluated. We will discuss the computation of $f'(\lambda)$ in the next part.
It is now clear that, given $\lambda^*$, the optimal policy for problem P1(c) (hence P1(a)) is the following:

\textit{ Let $a_i(X,\lambda)$ and $r_i(X,\lambda)$ be the active and passive functions associated with the  $i$-th single bandit problem, as defined in \eqref{eq:actFun} and \eqref{eq:pasFun}. Then at each time $t$, the $i$-th remote DTMC is selected to update if and only if ${a_i}({X_i}(t),{\lambda ^*}) \le {r_i}({X_i}(t),{\lambda ^*})$.	
}

We refer to the above policy as the optimal relaxed (OR) policy. Since problem P1(a) is a relaxation of the original RMAB problem P1, it is natural to construct a policy for P1 by rounding up the optimal policy of P1(a). The difference is that P1(a) does not require the number of transmitting processes in each time slot to be exactly $m$, while P1 does.

\subsection{The Gain Index Policy }

We propose the following index policy for the original RMAB problem P1:
\begin{definition}[Gain Index Policy]
	For each DTMC $i\in [M]$, define for each belief state of $J_i(\lambda^*)$, say $X=[x_1,x_2,\cdots,x_N]$, a gain index as follows:
	\begin{align} \label{eq:gdInd}
	{W_i}(X) = {\rho _i}\left[ {{V_i}({{\bf{T}}^{(i)}}X,{\lambda ^*}) - \mathop \sum \limits_{k = 1}^N {x_k}{V_i}({\bf{T}}_k^{(i)},{\lambda ^*})} \right].
	\end{align}
	At each time $t$, the $i$-th DTMC is assigned with a gain index $W_i(X_i(t))$, where $X_i(t)$  is the belief state of the associated bandit at this time. Then at each time, the gain index policy selects the $m$  DTMCs with the largest $m$  indices to transmit.
\end{definition}

According to Lemma 1, $V_i(X,\lambda)$ is a concave function of $X$ for any fixed $\lambda$; hence the gain indices are always non-negative. Let us justify the gain index policy by comparing it with the OR policy. Consider an arbitrary time slot $t$  and assume that the OR policy selects $y_t$  DTMCs to transmit at this time. Without loss of generality, suppose that the first $y_t$  DTMCs are activated by the OR policy. That is
\begin{align*}
\begin{cases}
{a_i}({X_i}(t),{\lambda ^*}) \le {r_i}({X_i}(t),{\lambda ^*}),\quad &1 \le i \le {y_t}\\
{a_i}({X_i}(t),{\lambda ^*}) > {r_i}({X_i}(t),{\lambda ^*}),\quad &{y_t} < i \le M,
\end{cases} 
\end{align*}
which implies that
\begin{align} \label{eq:dit}
{d_i}(t) \triangleq {r_i}({X_i}(t),{\lambda ^*}) - {a_i}({X_i}(t),{\lambda ^*}) =  \begin{cases}
\beta {W_i}({X_i}(t)) - {\lambda ^*} \ge 0,\quad &1 \le i \le {y_t}\\
\beta {W_i}({X_i}(t)) - {\lambda ^*} < 0,\quad &{y_t} < i \le M.
\end{cases} 
\end{align}
We can interpret $d_i(t)$  as the gain of applying active instead of passive action to the  $i$-th bandit at time $t$, i.e., the reduction of UoI for selecting the  $i$-th DTMC to transmit. From \eqref{eq:dit}, it is clear that $d_i(t)>d_j(t)$ if and only if ${W_i}({X_i}(t)) > {W_j}({X_j}(t))$. If $y_t = m$, then \eqref{eq:dit} means that exactly the first $m$ DTMCs have the largest $m$ indices; hence the gain index policy selects the same DTMCs as the OR policy. If $y_t\neq m$, then the gain index policy operates greedily in terms of $d_i(t)$. That is, it prefers the DTMC with a larger gain of transmitting. Consequently, a DTMC with a larger gain index value is preferred. 

The remaining issue is how to compute the gain indices efficiently. We first discuss how to compute $f'(\lambda)$  in the gradient method so that $\lambda^*$ can be determined. For this purpose, we define an auxiliary MDP for each DTMC, which is useful for computing $\partial {V_i}/\partial \lambda $.

\begin{definition}[Auxiliary MDP] \label{def:auxMDP}
	The auxiliary MDP of the  $i$-th DTMC, denoted by  $\mathcal{M}_i$, is defined as follows: it has the same state space, action space, transition rule, and discount factor as $J_i(\lambda)$; while the cost for state-action pair $(X,u)$ is $u$.
\end{definition}

Recall that $J_i(\lambda)$  denotes the belief MDP associated with the  $i$-th DTMC. Since  $J_i(\lambda)$ and $\mathcal{M}_i$  have the same state space and action space, any policy of  $J_i(\lambda)$ is also applicable to  $\mathcal{M}_i$. For an arbitrary policy $\pi$ of $J_i(\lambda)$, define
\begin{align} 
{h_i}(\pi ,X) \buildrel \Delta \over = {E_\pi }\left[ {\sum\limits_{t = 1}^\infty  {{\beta ^{t - 1}}{u_i}(t)|X} } \right].
\end{align}
Then $h_i(\pi,X)$ is the value function of $\mathcal{M}_i$ under policy $\pi$. Denote by $\pi_{i,\lambda}$ an optimal policy for $J_i(\lambda)$, and ${{\bf{P}}_{i,\lambda }}$ the transition matrix associated with $\pi_{i,\lambda}$. The cost function of $\mathcal{M}_i$ under policy $\pi_{i,\lambda}$ is given by
\begin{align}
{C_{i,\lambda }}(X) = \begin{cases}
1,{\text{ if }}{\pi _{i,\lambda }}{\text{ takes active action in state }}X\\
0,{\text{ if }}{\pi _{i,\lambda }}{\text{ takes passive action in state }}X.
\end{cases} 
\end{align}
Let $h_i(\pi)$ denote the vector form of $h_i(\pi,X)$ over all possible $X$. Then $h_i(\pi_{i,\lambda})$  can be computed by solving the following linear equation (i.e., policy evaluation): 
\begin{align} \label{eq:comp-hi}
h_i(\pi_{i,\lambda}) = {\left( {\mathbf{I} - \beta {{\bf{P}}_{i,\lambda }}} \right)^{ - 1}}{C_{i,\lambda }}.
\end{align}
We have shown in the proof of Theorem \ref{thm:Vlambda} that
\begin{align*}
\frac{{\partial {V_i}({\chi ^{(i)}},\lambda )}}{{\partial \lambda }} = {h_i}({\pi _{i,\lambda }},{\chi ^{(i)}}).
\end{align*}
Note that $J_i(\lambda)$ may have multiple optimal policies for some $\lambda$, and ${V_i}({\chi ^{(i)}},\lambda )$ is not differentiable in these $\lambda$. When this situation occurs in practical computations, we can choose  $\pi_{i,\lambda}$  to be an arbitrary optimal policy, and ${h_i}({\pi _{i,\lambda }},{\chi ^{(i)}})$ obtained by \eqref{eq:comp-hi} is the right or left derivative of ${V_i}({\chi ^{(i)}},\lambda )$.

Since the belief MDP of each single bandit problem has infinite states, directly computing the optimal policy by value iteration method or policy iteration method is computationally intractable. Thanks to the result established in Theorem \ref{thm:errorbound1}, we can approximate each single bandit problem using an  $L$-truncated belief MDP. It turns out that the approximation is usually good enough with moderate $L$.
In this case, computing the optimal policy of a single bandit problem and then solving \eqref{eq:comp-hi} to obtain  $\partial {V_i}/\partial \lambda $ are tractable. Note that the auxiliary MDP used to compute $\partial {V_i}/\partial \lambda $  is also a tractable  $L$-truncated version. Algorithm \ref{al:CWI} summarizes the whole procedure of computing the gain indices.

\begin{algorithm}[!ht]
	\caption{Computing the Gain Indices }
	\label{al:CWI}
	%\LinesNumbered
	\KwIn{$M,\{ {{\bf{T}}^{(i)}},i \in [M]\} $, discount factor $\beta$.}
	\begin{itemize}[leftmargin=40pt]
		\item [\textbf{Step 1}:] Select truncation factor $L$ and stepsize factor $c$, specify $\epsilon>0$, and set $n=0,\lambda_0=0$. \\
		Let $\pi_{i,\lambda_0}$ be the policy that takes active action in all belief states, $i\in [M]$.
		
		\item [\textbf{Step 2}:] \textbf{(Computing the derivative}) \\
		\textbf{For} $i=1 $ to $M$ \\
		\quad Determine the  $L$-truncated belief MDP $J_i^L({\lambda _n})$.\\
		\quad ${\pi _{i,{\lambda _n}}} \leftarrow $ \textbf{PolicyIteration}($J_i^L({\lambda _n})$,${\pi _{i,{\lambda _{n - 1}}}}$). //skip this line when $n=0$ \\
		  \quad Compute ${h_i}({\pi _{i,{\lambda _n}}})$ by \eqref{eq:comp-hi}. \\
		\textbf{End} \\
		Compute the derivative of $f(\cdot)$: $f'({\lambda _n}) = \sum\limits_{i = 1}^M {{h_i}\left( {{\pi _{i,{\lambda _n}}},{\chi ^{(i)}}} \right)}  - {m}/(1 - \beta)$.
		
		\item[\textbf{Step 3}:] ${\lambda _{n + 1}} = {\lambda _n} + \frac{c}{n}f'({\lambda _n})$. \\
		If $f'({\lambda _n})f'({\lambda _{n - 1}}) \le 0{\text{ and }}|{\lambda _n} - {\lambda _{n - 1}}| < \epsilon$, let ${\lambda ^*} = \min \{ {\lambda _n},{\lambda _{n - 1}}\} $ and go to step 4;\\
		otherwise, increment $n$ by 1 and return to step 2.
		
		\item[\textbf{Step 4}:] (\textbf{Computing the gain indices}) \\
		\textbf{For} $i=1 $ to $M$ \\
		\quad Policy evaluation: ${V_i} = {(\mathbf{I} - \beta {{\bf{P}}_{i,{\lambda ^*}}})^{ - 1}}(H + {C_{i,{\lambda ^*}}})$.\\
		\quad Compute the gain indices by \eqref{eq:gdInd}. \\
		\textbf{End} 			
	\end{itemize}	
	\vspace{-4mm}
	\rule{16cm}{0.8pt}		
	Note: \textbf{PolicyIteration}($J,\pi$) refers to computing the optimal policy for MDP $J$  by the policy iteration algorithm with initial policy $\pi$.
\end{algorithm}

\section{UoI Scheduling with Average Cost Criterion} \label{sec:average}
This section studies UoI-scheduling with the average cost criterion, i.e., problem P2 given in \eqref{eq:P2}, with the aim of extending the gain index policy to this setting. As in Section III, we first relax and decouple the original RMAB problem into   single bandit problems. Then we establish the properties of the single bandit problem needed to develop the gain index policy. By doing so, the gain index policy for the average cost problem can then be developed.

First, we relax the constraint in P2 by
\begin{align} \label{eq:relaxStP2}
E\left[\mathop {\lim }\limits_{n \to \infty } \frac{1}{n}\sum\limits_{t = 1}^n {\sum\limits_{i = 1}^M {{u_i}(t)} }\right]  = m.
\end{align}
The above equation means that the number of bandits being activated in each slot, averaged over time, is $m$. Replacing constraint \eqref{eq:P1-st1} in P2 by \eqref{eq:relaxStP2} and converting the problem by the Lagrange multiplier method yield
\begin{align} \label{eq:P2a}
{\text{P2(a): }}\mathop {\min }\limits_{\{ {u_i}(t)\} } \mathop {\sup }\limits_{\lambda  \ge 0} {\rm{  E}}\left[ {\mathop {\lim }\limits_{n \to \infty } \frac{1}{n}\sum\limits_{t = 1}^n {\sum\limits_{i = 1}^M {\left[ {H\left( {{X_i}(t)} \right) + \lambda {u_i}(t)} \right]} |\chi } } \right] - m\lambda .
\end{align}
Applying the similar argument as in the treatment of P1(a), we can interchange the min and sup in P2(a), leading to an equivalent problem:
\begin{align} \label{eq:P2b}
{\text{P2(b): }}\mathop {\sup }\limits_{\lambda  \ge 0} \mathop {\min }\limits_{\{ {u_i}(t)\} } {\rm{  E}}\left[ {\mathop {\lim }\limits_{n \to \infty } \frac{1}{n}\sum\limits_{t = 1}^n {\sum\limits_{i = 1}^M {\left[ {H\left( {{X_i}(t)} \right) + \lambda {u_i}(t)} \right]} |\chi } } \right] - m\lambda.
\end{align}
For a fixed  $\lambda$, problem P2(b) can be decoupled into $M$  single bandit problems:
\begin{align} \label{eq:Gilab}
{G_i}(\lambda ): = \mathop {\min }\limits_{\{ {u_i}(t)\} } {\rm{  E}}\left[ {\mathop {\lim }\limits_{n \to \infty } \frac{1}{n}\sum\limits_{t = 1}^n {\left[ {H\left( {{X_i}(t)} \right) + \lambda {u_i}(t)} \right]|{\chi ^{(i)}}} } \right],\quad i \in [M].
\end{align}
Note that $G_i(\lambda)$  and $J_i(\lambda)$  have the same belief MDP model but different objectives. We refer to $G_i(\lambda)$  as the single bandit problem with average cost criterion associated with the  $i$-th DTMC.

\subsection{The Single Bandit Problem} \label{subsec:ave-single}
This part investigates the single bandit problem given in \eqref{eq:Gilab}, which is a belief MDP with the average cost criterion. For simplicity, we drop the bandit index from all notations in this part. The belief MDP is the same as that in Section III.B, except that we are considering the average cost criterion here. 

Recall that the original belief state space of the single bandit problem is denoted by $\mathcal{S}$. For any $X\in \mathcal{S}$, let $g(X)$ denote the optimal average cost (i.e., minimum time-average UoI) with the initial state $X$. It is well-known that $g(X)$  may vary from different initial states if there exist policies that generate Markov chains consisting of multiple recurrent classes. This is the so-called multichain model. We refer readers interested in the unichain and multichain models to \cite{puterman1994markov} for their definitions. Fortunately, we have the following result for the single bandit problem $G(\lambda)$:
\begin{lemma} \label{Lem:g-constant}
	Given any $\lambda \ge 0$ and $\rho \in (0,1]$, $g(X)$ is a constant function independent of $X$ for all $X\in \mathcal{S}$.
\end{lemma}
\begin{IEEEproof}
	For $\rho \in (0,1)$, it is easy to see from the transition rule \eqref{eq:SbTrans} that the belief MDP is unichain. In this case $g(X)$ is a constant. While if $\rho = 1$, we can find some examples showing that the belief MDP is multichain. However, it turns out that the problem can always be optimized by a policy with a unichain transition structure. Therefore, given any $\lambda \ge 0$ and $\rho \in (0,1]$, there exists an optimal policy that generates a Markov chain consisting of a single recurrent class and some transient states. We thus conclude that the long-term average cost is invariant to the initial state. The detailed proof for the special case of  $\rho = 1$ is omitted because it is tedious and lengthy.
\end{IEEEproof}

Let ${\Gamma}$  denote the set of policies resulting in an average cost that is a constant function of initial states. Lemma \ref{Lem:g-constant} implies that we can focus on policies in  ${\Gamma}$ and treat the single bandit problem $G(\lambda)$ as a unichain MDP. Therefore, the optimal policy can be determined by the Bellman equation \cite{puterman1994markov}.
That is, for any $\lambda\ge 0$ and $X\in \mathcal{S}$,
\begin{align}
Z(X,\lambda ) + g(X,\lambda ) = H(X) + \min \left\{ {\lambda  + \rho \mathop \sum \limits_{i = 1}^N {x_i}Z({{\bf{T}}_i},\lambda ) + (1 - \rho )Z({\bf{T}}X,\lambda ),Z({\bf{T}}X,\lambda )} \right\}.
\end{align}
where $Z(X,\lambda)$ is the differential value function. As in the discounted cost problem, we express $Z$  and $g$  as bivariate functions of $X$  and $\lambda$  to emphasize the dependency on $\lambda$. Without loss of generality, we let $Z({{\bf{T}}_1},\lambda ) = 0$ for all $\lambda\ge 0$. Since $g(X,\lambda)$ is invariant over  $X$ for any given $\lambda$, we will simply write it as $g(\lambda)$  whenever there is no ambiguity. That is, let $g(\lambda) = g(X,\lambda)$ for all $X\in \mathcal{S}$.

For any policy $\pi \in \Gamma$, denote by $g_\pi(\lambda)$ the long-term average cost incurred by policy $\pi$, ${P_\pi(X|Y)}$  the transition probability from belief state  $X$ to belief state $Y$  under policy $\pi$, and $\pi(X)\in \{0,1\}$ the action taken in state $X$ by policy $\pi$. The lemma below is useful.

\begin{lemma} \label{Lem:g-pro}
	For the single bandit problem $G(\lambda)$ with $\lambda \in[0,\infty)$,
	\begin{itemize}
		\item [1.] Given an arbitrary $\pi \in \Gamma$, the derivative of $g_\pi(\lambda)$, denoted by $g'_\pi(\lambda)$, is a constant function of $\lambda$  and can be computed by solving the following equations:
		\begin{align*}
		z(X) + {g'_\pi }(\lambda ) = \pi (X) + \sum\limits_{Y \in {\cal S}} {{P_\pi }(Y|X)z(Y)} ,\quad X \in {\cal S},
		\end{align*}
		where we can let $z({{\bf{T}}_1} ) = 0$   and determine $g'_\pi(\lambda)$  and other $z(X)$  by the above equations. 
		\item [2.]	The optimal average cost  $g(\lambda)$ is an increasing, concave, and piecewise linear function of $\lambda$.
		
		\item[3.] For any fixed $\lambda$, let $\Pi_\lambda$ denote the set of optimal policies for $G(\lambda)$. If $\Pi_\lambda = \{\pi_\lambda\}$ is a singleton, then $g(\lambda)$ is differentiable at $\lambda$  and $g'(\lambda ) = {g'_{{\pi _\lambda }}}(\lambda )$. Otherwise, $g(\lambda)$  may be non-differentiable at $\lambda$  and its left and right derivatives are given by
		\begin{align*}
		g'({\lambda _ - }) = \mathop {\max }\limits_{\pi  \in {\Pi _\lambda }} {g'_\pi }(\lambda ),\quad g'({\lambda _ + }) = \mathop {\min }\limits_{\pi  \in {\Pi _\lambda }} {g'_\pi }(\lambda ).
		\end{align*}
	\end{itemize}
\end{lemma}
\begin{IEEEproof}
	See Appendix III.
\end{IEEEproof}

Statement 1 of Lemma \ref{Lem:g-pro} shows that, for any $\pi \in\Gamma$, $g'_\pi(\lambda)$ is the average cost of the auxiliary MDP under policy $\pi$. This result is useful in computing $g'(\lambda)$. The role of the following results to $G(\lambda)$  is the same as Corollary 1 to  $J(\lambda)$. 

\begin{lemma} \label{Lem:g-coro}
	For the single bandit problem $G(\lambda)$,
	\begin{itemize}
		\item [1.] If $\lambda = 0$, the optimal policy is to take active action in all belief states and $g'({0_ + }) = 1$.
		
		\item[2.] There exists an $\bar{\lambda}$ such that for any $\lambda > \bar{\lambda}$, the optimal policy is to take passive action in all belief states and $g'(\lambda ) = 0$.
	\end{itemize}
\end{lemma}
\begin{IEEEproof}
	This lemma can be proved by a similar method as in proving Corollary 1. Namely, we can extend Lemma 1 and Lemma 2 to the setting of average cost criterion, and then use them to verify the above two statements. But this method is a little bit lengthy. Hence we provide an alternative method, which uses the asymptotic relationship between the discounted cost criterion and the long-term average cost criterion. See Appendix III.
\end{IEEEproof}

Just like in the discounted cost problem, we can approximate $G(\lambda)$  using an  $L$-truncated MDP. Denote by $G^L(\lambda)$ the $L$-truncated MDP of $G(\lambda)$. The construction of  $G^L(\lambda)$ is the same as $J^L(\lambda)$ , except that  $G^L(\lambda)$ is with the average cost criterion. The optimal average cost of $G^L(\lambda)$  is denoted by  $g^L(\lambda)$. The following theorem provides a bound for the approximation error in terms of the average cost.
\begin{theorem} \label{Thm:bound-ave}
	Let $L$ be a positive integer such that
	\begin{align*}
	\mathop {\max }\limits_{i \in [N],k \ge 0} \left| {H({\bf{T}}_i^{L + k}) - H(\omega )} \right| \le {\sigma _L},
	\end{align*}
	where $\sigma_L > 0$. Then for any $\lambda\in[0,\infty)$ and $X\in \mathcal{S}^L$,
	\begin{align*}
	\left| {g(\lambda ) - {g^L}(\lambda )} \right| \le {\sigma _L}.
	\end{align*}
\end{theorem}
\begin{IEEEproof}
	See Appendix III.
\end{IEEEproof}

\subsection{The Gain Index Policy}
We have proved the desired properties of the single bandit problem under the average cost criterion. Next, let us return to problem P2(b) and derive the gain index policy for problem P2. 

For $i\in [M]$ and $\lambda\ge 0$, denote by $g_i(\lambda)$ the optimal average cost of the single bandit problem associated with the  $i$-th DTMC (i.e., $G_i(\lambda)$). Define a function for $\lambda \in [0,\infty)$:
\begin{align*}
l(\lambda ) = \sum\limits_{i = 1}^M {{g_i}(\lambda )}  - m\lambda .
\end{align*}
Then problem P2(b) can be written as
\begin{align}
{\text{P2(c): }}\mathop {\sup }\limits_{\lambda  \ge 0} {\rm{ }}l(\lambda ).
\end{align}
According to Lemma \ref{Lem:g-pro}, $l(\lambda)$  is an increasing, concave, and piecewise linear function of  $\lambda$. Lemma \ref{Lem:g-coro} implies that
\begin{align*}
\mathop {\max }\limits_{\lambda  \ge 0} l'(\lambda ) &= l'({0_ + }) = M - m > 0,\\
\mathop {\min }\limits_{\lambda  \ge 0} l'(\lambda ) &= \mathop {\lim }\limits_{\lambda  \to \infty } l'(\lambda ) =  - m < 0.
\end{align*}
We thus can use the same gradient method as in problem P1(c) to solve P2(c). That is, using the gradient iteration \eqref{eq:gradient} and the stopping criterion \eqref{eq:stopCrit} yields a  $\epsilon$-optimal solution to P2(c), just replacing  $f'(\lambda)$ with $l'(\lambda)$. Theorem \ref{Thm:finitestop} is also applicable to this problem. Therefore, given any  $\epsilon>0$, the gradient method will stop within a finite number of iterations. 

Using the same argument as in the RMAB with the discounted cost criterion, we can obtain the gain index policy for the RMAB problem P2 by rounding up the optimal policy for P2(c). In particular, the gain indices for each bandit are defined below.
\begin{definition}[Index for the Average Cost Problem]
	Denote by $\lambda^a$ an optimal solution to P2(c). For each remote DTMC  $i\in[M]$, define for each belief state of $G_i(\lambda^a)$, say $X = [{x_1},{x_2}, \cdots ,{x_N}]$, a gain index as follows:
	\begin{align}
	{W_i}(X) = {\rho _i}\left[ {{Z_i}({{\bf{T}}^{(i)}}X,{\lambda ^a}) - \mathop \sum \limits_{k = 1}^N {x_k}{Z_i}({\bf{T}}_k^{(i)},{\lambda ^a})} \right],
	\end{align}
	where $Z_i(\cdot,\lambda^a)$ is the differential value function of  $G_i(\lambda^a)$.
\end{definition}

After computing the gain indices, the gain index policy selects at each time the $m$  DTMCs with the largest  $m$ indices to update their states. The algorithm to compute the gain indices is similar to Algorithm 1. 
A minor modification is needed in step 2---we compute ${g'_i}({\lambda _n})$, instead of ${h_i}({\pi _{i,{\lambda _n}}})$, by Lemma \ref{Lem:g-pro}, and then evaluate $l'(\cdot)$ by
\begin{align*}
l'(\lambda ) = \sum\limits_{i = 1}^M {{g'_i}(\lambda )}  - m.
\end{align*}
In addition, \eqref{eq:PE-ave} should be used for policy evaluation in step 4.
Theorem \ref{Thm:bound-ave} suggests that an  $L$-truncated MDP can be used to approximate each single bandit problem. 

\subsection{Discussion}
We have developed a gain index policy for UoI-scheduling with discounted and average cost criteria. The policy is justified under the RMAB framework. Remarkably, we find that the gain index policy has great universality in the sense that it applies to general RMABs with bounded cost functions. To see this, we review the properties needed to establish the gain index policy and show that the only requirement is the boundedness of the cost function. We also have a discussion on the computation complexity of the policy at the end of this part.

Let us first consider a specific class of RMABs that includes the UoI scheduling problem as a special case. Formally, 

\begin{definition}[C-type RMAB]
	An RMAB with the form of P1 (discounted cost criterion) or P2 (average cost criterion) is said to be C-type if the cost function $H(X)$  is a bounded concave function of the belief state $X\in \Omega$.
\end{definition}

We remind the reader that boundedness and concavity are all the  properties required for the cost function $H(X)$  throughout our derivation; hence all the results established in this paper remain valid as long as $H(X)$  is a bounded concave function of the belief state (not limited to UoI). Therefore, the gain index policy proposed in this paper can be applied to any C-type RMAB.

In fact, with a minor modification, our method is viable for more general RMABs. Let $\Xi  = \left( {\{ {{\cal X}_i}\} _{i = 1}^N,\{ {{\cal P}_i}\} _{i = 1}^N,\{ {F_i}\} _{i = 1}^N,\beta } \right)$ denote an RMAB with $N$  bandits, where ${{\cal X}_i},{{\cal P}_i},{F_i}$ denote the state space, transition matrix, and cost function of the  $i$-th bandit, respectively. We assume that $F_i(\cdot)$  is a bounded function on $\mathcal{X}_i$. Further, $\beta=1$  indicates that the RMAB is of average cost criterion. While if  $\beta\in[0,1)$, then the RMAB is of discounted cost criterion with discount factor $\beta$. We next show that the gain index policy is applicable to $\Xi $ with $\beta\in[0,1)$. 

Note that the relaxation and decomposition treatments given by \eqref{eq:RelSt}-\eqref{eq:Ji} are valid for $\Xi$. The next key step is transforming problem P1(b) to P1(c)---maximizing a concave and piecewise linear function $f(\lambda)$. The concavity and piecewise linearity established in Theorem \ref{thm:Vlambda} are essential for the gain index policy because they allow us to compute the optimal  $\lambda$ efficiently by the gradient method. Without the concavity of $f(\lambda)$, it may be difficult to determine $\lambda^*$, and the gain indices can not be computed as a result. Fortunately, it is easy to verify that Theorem \ref{thm:Vlambda} is valid for the single bandit problems of $\Xi$. Therefore, the relaxed problem of $\Xi$  can be transformed into the form of P1(c), in which $f(\lambda)$  is still concave and piecewise linear. Then P1(c) can be easily solved because of the concavity and piecewise linearity of $f(\lambda)$. Without the assumption that $F_i(X)$ is concave, Lemma \ref{Lem:Vxconcave} and Lemma \ref{Lem:ConvexAct} are not valid in general. Hence statement 1 of Corollary \ref{Coro:lambda0} may not hold either. As a result, it is possible that $f'({0_ + }) < 0$. If $f'({0_ + }) < 0$, then we can directly determine that $\lambda^* = 0$. Otherwise, Lemma \ref{Lem:f'} and Theorem \ref{Thm:finitestop} are still applicable; we thus can use the gradient method to compute $\lambda^*$. For the  $i$-th single bandit problem with service charge $\lambda^*$, its optimal policy can be determined by the following Bellman equation:
\begin{align*}
{V_i}(X,{\lambda ^*}) = {F_i}(X) + \min \left\{ {\lambda^*  + \beta \sum\limits_{Y \in {\cal X}} {{{\cal P}_i}(Y|X,1){V_i}(Y,{\lambda ^*})} ,\beta \sum\limits_{Y \in {\cal X}} {{{\cal P}_i}(Y|X,0){V_i}(Y,{\lambda ^*})} } \right\},
\end{align*}
where ${{\cal P}_i}(Y|X,u)$ is the transition probability from state $X$  to state $Y$  under action  $u$.
Using the same argument as in Section \ref{sec:policy}, we can develop the gain index policy for $\Xi$. As discussed around \eqref{eq:dit}, the gain index should reflect the gain of applying active instead of passive action to the associated bandit. Therefore, the gain index of $X \in {\cal X}{_i}$  is given by
\begin{align} \label{eq:Wgeneral}
{W_i}(X) = \sum\limits_{Y \in {\cal X}} {{{\cal P}_i}(Y|X,0){V_i}(Y,{\lambda ^*})}  - \sum\limits_{Y \in {\cal X}} {{{\cal P}_i}(Y|X,1){V_i}(Y,{\lambda ^*})}.
\end{align}
Note that \eqref{eq:Wgeneral} reduces to \eqref{eq:gdInd} when $\Xi$ is C-type (to see this, substituting the transition probability given by \eqref{eq:SbTrans} into the above equation). Using the same argument can verify that the gain index policy is also applicable to $\Xi$ with $\beta = 1$. The gain index for the average cost problem is also of the form of \eqref{eq:Wgeneral}, just replace $V_i(X,\lambda)$ by the differential value function. We do not repeat the discussion for simplicity.

The universality of the gain index policy makes it a promising method for general RMABs. Compared with the celebrated Whittle index policy, an advantage of our gain index policy is that it does not require establishing “indexability”. It is well-known that the Whittle index policy is not naturally applicable to all RMABs because not all RMABs are indexable. In fact, proving the indexability may be very difficult for some RMABs (e.g., the UoI-scheduling problem studied in this paper). Even if the indexability can be established, computing the Whittle indices is usually hard because of the lack of closed-form expression. In contrast, the gain index always exists and has a clear expression. As will be demonstrated by simulations in the next section,  
our gain index policy is a promising alternative method for RMABs: it can be used when the Whittle index policy is not viable and it performs as well as the Whittle index policy even when the Whittle index policy is viable.

Concerning computation complexity, the computation of gain indices needs to determine the optimal $\lambda$  by the gradient method; computing the gradient involves policy iteration for every single bandit problem. But it turns out that the computation complexity of our method is acceptable as long as $N$  (the dimension of the remote DTMCs) is not too large. 
First, the concavity and piecewise linearity of the objective function allow our algorithm to determine the optimal $\lambda$  efficiently if a proper stepsize sequence is used. 
Second, given an arbitrary $\lambda$, we have decoupled the RMAB into $M$  single bandit problems. Policy iteration is carried out for the single bandit problems instead of the original RMAB; hence the complexity of our method is linearly increasing with $M$  (it is exponentially increasing in the original RMAB). That is, the complexity is already significantly reduced.
Third, there is a trick in Algorithm 1 that speeds up the computation. For each truncated belief MDP $J_i^L(\lambda )$, we use the optimal policy of $J_i^L(\lambda_n )$  to initialize the policy iteration for  $J_i^L(\lambda_{n+1} )$. It is well-known that the policy iteration algorithm converges quickly if the initial policy is close to the optimal policy \cite{puterman1994markov}. The value functions of $J_i^L(\lambda_{n} )$  and  $J_i^L(\lambda_{n+1} )$ are close if $|{\lambda _n} - {\lambda _{n + 1}}|$  is small, and so are their optimal policies. This becomes true as $\{\lambda_n\}$  converges to the optimal $\lambda$. 
Finally, our method only requires offline computation. That is, the gain indices of all bandits can be pre-computed before real-time scheduling. Once the offline computation is completed, the results can be incorporated into a lookup table for use in real-time execution. Hence it is deployable in practical systems, even with a certain complexity for the offline computation.

{
\section{Asymptotic Optimality}  \label{sec:asy-opt}
This section establishes the asymptotic optimality of the gain index policy. In particular, we show that the per-bandit value of the gain index policy converges to the optimal per-bandit value as $m$ and $M$ tend to $\infty$ with $\alpha=m/M$ fixed. We first prove that the gain index policy is asymptotically optimal for the discounted cost problem. The results can then be easily extended to the average cost problem.

Throughout this section, we assume that the bandits of the RMAB can be divided into $Q$ classes, where $Q$ is a finite positive integer. The bandits of the same class are stochastically identical (i.e., the associated DTMCs share an identical transition matrix). Let $\mathcal{Q}_k\in [M]$ represent the $k$-th class, and thus $\cup^Q_{k=1}\mathcal{Q}_k=[M]$. Further, denote by $q_k\in (0,1]$ the proportion of the $k$-th class, then $\sum_{k=1}^Q q_k = 1$ and the $k$-th class consists of $|\mathcal{Q}_k|=Mq_k$ bandits. We consider $Q$ and $\{q_k:k\in [Q] \}$ to be arbitrary but fixed.

\subsection{The Discounted Cost Problem}
We first focus on the discounted cost problem. The asymptotic optimality is established by comparing the gain index policy with the OR policy. Hence it is necessary to first remark on the properties of the OR policy. As stated in Section \ref{subsec:Dpolicy-A}, the OR policy is defined independently for each bandit. That is, under the OR policy, the actions of a bandit are independent of other bandits. Let $\pi^*$ be an optimal policy for the relaxed problem (i.e., the discounted cost objective subject to the relaxed constraint \eqref{eq:RelSt}). Then $\pi^*$ satisfies the relaxed constraint:
\begin{align} \label{eq:pi*}
E_{\pi^*}\left[ {\sum\limits_{t = 1}^\infty  {{\beta ^{t - 1}}} \mathop \sum \limits_{i = 1}^M {u_i}(t)} \right] =  \frac{m}{{1 - \beta }}.
\end{align}
According to the duality theory, $(\lambda^*,\pi^*)$ is an optimal solution to the sup-min problem P1(b). Note that there may exist multiple policies that are equally optimal for problem P1(b) (given $\lambda^*$, a single bandit problem may have multiple optimal stationary policies). In this part, the OR policy exclusively refers to $\pi^*$.

Let ${\cal W} \buildrel \Delta \over = \prod_{i = 1}^M {{{\cal S}^{(i)}}} $ denote the state space of the RMAB and $\zeta:\mathcal{W} \to [0,1] $ denote the equilibrium distribution of the RMAB governed by the OR policy. We make the following assumption:
\begin{assumption}
	The initial state of the RMAB follows the equilibrium distribution $\zeta$.
\end{assumption}
  
We need this assumption because the behavior of the OR policy becomes stable (in the probabilistic sense) after entering the steady state. With this assumption, the state distribution of the RMAB governed by the OR policy is $\zeta$  at any time. Under Assumption 1, denote by $J^{opt}_M$ the optimal value of the RMAB (i.e., problem P1), $J^{rel}_M$ the optimal value of the relaxed problem, and $J^{ind}_M$  the value of the gain index policy. The subscript $M$ in these notations represents the total number of bandits in the RMAB. It is clear that, for any positive integer $M$, 
\begin{align*}
J_M^{ind} \ge J_M^{opt} \ge J_M^{rel}.
\end{align*}
Essentially, we aim to show that in the steady state, the OR policy behaves almost the same as the gain index policy. This is formally presented in Proposition 1. On this basis, we prove the asymptotic optimality by showing $J_M^{ind}/M \to J_M^{rel}/M$ as $M\to \infty$. Before that, we need the following lemma.

\begin{lemma} \label{lem:7}
	Suppose Assumption 1 holds. For the RMAB governed by the OR policy, there exists an $\alpha_k\in [0,1]$ for the $k$-th class such that for any $i\in \mathcal{Q}_k$, 
	\begin{align*}
	\Pr \{ {u_i}(t) = 1\}  = \Pr \left\{ {{a_i}\left( {{X_i}(t),{\lambda ^*}} \right) \le {r_i}\left( {{X_i}(t),{\lambda ^*}} \right)} \right\} = {\alpha _k},\quad \forall t.
	\end{align*}
	Moreover, $\sum_{k=1}^Q q_k \alpha_k =\alpha\ $.
\end{lemma}
\begin{IEEEproof}
	See Appendix IV.
\end{IEEEproof}

Let $y_t^{(M)} = \sum\nolimits_{i = 1}^M {{u_i}(t)} $ denote the number of bandits taking active action at time $t$  under the OR policy. Then $\{ y_t^{(M)}:t \ge 1\} $ is a stochastic process, and its distribution is related to the state distribution  of the RMAB. Note that the OR policy is stationary. At an arbitrary time, the probability of taking active action for a bandit is essentially the probability of visiting a state where the active action is optimal for that bandit; hence it is determined by the state distribution at the given time. In general, the state distribution becomes stable only after the system enters the steady state. Hence we can not expect the OR policy to be close to the gain index policy before entering the steady state. This explains why we need Assumption 1.
\begin{proposition} \label{pro:1}
	Suppose Assumption 1 holds. Then for any positive integer $k$,
	\begin{align*}
	\lim\limits_{M\to \infty}\Pr \left\{ {\left| {\frac{{y_t^{(M)}}}{M} - \frac{m}{M} } \right| \le \frac{k}{{\sqrt {4M} }}} \right\} \ge \Phi (k) - \Phi ( - k),\quad \forall t
	\end{align*}
	where $\Phi (k)$  is the cumulative distribution function of the standard normal distribution.
\end{proposition}
\begin{IEEEproof}
	See Appendix IV.
\end{IEEEproof}

It is well known that $\Phi (k) - \Phi ( - k)$ sharply tends to 1 as $k$ increases. Hence Proposition \ref{pro:1} means that $y_t^{(M)}/M \to \alpha $ with a probability arbitrarily close to 1. As we discussed around \eqref{eq:dit}, the OR policy and the gain index policy select the same bandits at time $t$ if $y_t^{(M)} = m$. If $y_t^{(M)} \neq m$, then the two policies take different actions for exactly $|y_t^{(M)} - m|$ bandits. Hence Proposition \ref{pro:1} implies that, in the steady state, the OR policy behaves almost the same as the gain index policy. With this result, it is natural to conjecture that $J_M^{ind}/M \to J_M^{rel}/M$ as $M\to \infty$. We formally establish this result in the following theorem.
\begin{theorem} \label{thm:optimality}
	Suppose Assumption 1 holds. Then
	\begin{align*}
	\mathop {\lim }\limits_{M \to \infty } \frac{1}{M}J_M^{ind} = \mathop {\lim }\limits_{M \to \infty } \frac{1}{M}J_M^{opt} = \mathop {\lim }\limits_{M \to \infty } \frac{1}{M}J_M^{rel}.
	\end{align*}
\end{theorem}
\begin{IEEEproof}
	See Appendix IV.
\end{IEEEproof}

\subsection{The Average Cost Problem}
In the average cost problem, the asymptotic optimality of the gain index policy holds without Assumption 1. In fact, the technique of proving the asymptotic optimality in this setting is similar to that used in the discounted cost problem. The focus of this part is to state the results rigorously and point out why Assumption 1 is not required in this setting.

Like before, we first remark on the OR policy for the average cost problem. Let $\pi^a$ denote an optimal policy for the relaxed average cost problem (i.e., P2 relaxed by constraint \eqref{eq:relaxStP2}). Then we have
\begin{align} \label{eq:pi-a}
 {E_{{\pi ^a}}}\left[\mathop {\lim }\limits_{n \to \infty } {\frac{1}{n}\sum\limits_{i = 1}^M {\sum\limits_{t = 1}^n {{u_i}(t)} } } \right] = m.
\end{align}
Given that $\lambda^a$  is an optimal solution to problem P2(c), $(\lambda^a,\pi^a)$ is an optimal solution to problem P2(b). Just like \eqref{eq:pi*} being used to prove Lemma \ref{lem:7} in the discounted cost problem, we can use \eqref{eq:pi-a} similarly to show that Lemma \ref{lem:7} also holds for the average cost problem. The only difference is that the OR policy in this setting refers to $\pi^a$. It follows immediately that Proposition \ref{pro:1} is also valid for this setting. Proposition \ref{pro:1} indicates that the OR policy behaves almost like the gain index policy in the steady state. The RMAB governed by the OR policy could enter the steady state within a finite time. Therefore, as far as the long-term average cost is concerned, the difference between the two policies in the transient state is negligible. This is why Assumption 1 is not needed in this setting. The explicit reason in the technical view will be clarified in the proof of Theorem 6.

Denote by $G^{opt}_M$ the optimal average cost of the RMAB (i.e., problem P2), $G^{rel}_M$ the optimal average cost of the relaxed RMAB (relaxed by constraint \eqref{eq:relaxStP2}), and  $G^{ind}_M$ the average cost of the gain index policy. For any positive integer $M$, we have
\begin{align*}
G_M^{ind} \ge G_M^{opt} \ge G_M^{rel}.
\end{align*}
In the following theorem, we prove the asymptotic optimality of the gain index policy by showing that $(G_M^{ind} - G_M^{rel})/M \to 0$ as $M\to \infty$. 
\begin{theorem} \label{thm:ave-opt}
	The gain index policy is asymptotically optimal for the RMAB with average cost criterion:
	\begin{align*}
	\mathop {\lim }\limits_{M \to \infty } \frac{1}{M}G_M^{ind} = \mathop {\lim }\limits_{M \to \infty } \frac{1}{M}G_M^{opt} = \mathop {\lim }\limits_{M \to \infty } \frac{1}{M}G_M^{rel}.
	\end{align*}
\end{theorem}
\begin{IEEEproof}
	The main idea of the proof is similar to that of Theorem \ref{thm:optimality}. We thus provide a brief proof with a main focus on addressing the technical issues caused by the average cost criterion, see Appendix IV.
\end{IEEEproof}

The asymptotic optimality implies that the gain index policy is near optimal when $M$ is large. This property, plus the low computation complexity brought by decoupling the RMAB into $M$ single bandit problems during the computing procedure, make the gain index policy an excellent algorithm for large-scale RMABs.

}

\section{Simulation} \label{sec:sim}
This section presents simulation results to demonstrate the performance of the proposed index policy. We first compare the gain index policy with the optimal policy in the setting of $M=2$ (i.e., 2 remote DTMCs) --- the optimal policy can be computed for this simple case. Then, we consider the setting of multiple DTMCs and provide the results of the Whittle index policy and two other policies for benchmarking purposes. Finally, we demonstrate that the gain index policy performs well in general RMABs. 
Throughout this section, the results of the optimal policy are found using value iteration and relative value iteration for the discounted and average cost criteria, respectively. For other policies, we organized the simulation runs into groups, with each group containing $50$ independent runs. Each simulation run lasts for $10^5$  slots. Then the performance of each policy is evaluated by the average result of the $50$ independent runs in a group.

Table \ref{tab:M2N3} presents the results of the optimal policy and the gain index policy for the UoI-scheduling problem with 2 remote DTMCs ($M=2$) and 1 communication channel ($m=1$). Each DTMC has 3 states ($N=3$). For the settings of the discounted cost criterion, the discount factor $\beta=0.9$.
In addition, the communication channel is set to be reliable for both DTMCs (${\rho _1} = {\rho _2} = 1$). 
Table \ref{tab:M2N3} shows that the gain index policy obtains near-optimal performance in all groups under both criteria.

\begin{table}[h]
	\centering
	\caption{UoI of the gain index policy and the optimal policy for 2 DTMCs with 3 states.}
	\label{tab:M2N3}
	\begin{tabular}{|c|cc|cc|}
		\hline
		\multirow{2}{*}{Group} & \multicolumn{2}{c|}{The Discounted Cost Criterion}         & \multicolumn{2}{c|}{The Average Cost Criterion}          \\ \cline{2-5} 
		& \multicolumn{1}{c|}{Optimal Policy} & Gain Index Policy & \multicolumn{1}{c|}{Optimal Policy} & Gain Index Policy\\ \hline
		A1                     & \multicolumn{1}{c|}{27.31}   & 27.35  & \multicolumn{1}{c|}{2.71}    & 2.71   \\ \hline
		A2                     & \multicolumn{1}{c|}{25.17}   & 25.23  & \multicolumn{1}{c|}{2.47}    & 2.47   \\ \hline
		A3                     & \multicolumn{1}{c|}{24.28}   & 24.29  & \multicolumn{1}{c|}{2.358}   & 2.359  \\ \hline
	\end{tabular}
\end{table}

Table \ref{tab:M2N4} shows the results of the optimal policy and the gain index policy for the UoI-scheduling problem in the setting of $N = 4,M = 2,m = 1$. The communications are unreliable and ${\rho _1} = 0.7,{\rho _2} = 0.8$. The discount factor $\beta$  is set to be  $0.8$ for the discounted cost criterion. Again, the gain index policy achieves near-optimal performance in all settings.

\begin{table}[h]
	\centering
	\caption{UoI of the gain index policy and the optimal policy for 2 DTMCs with 4 states.}
	\label{tab:M2N4}
	\begin{tabular}{|c|cc|cc|}
		\hline
		\multirow{2}{*}{Group} & \multicolumn{2}{c|}{The Discounted Cost Criterion}         & \multicolumn{2}{c|}{The Average Cost Criterion}          \\ \cline{2-5} 
		& \multicolumn{1}{c|}{Optimal Policy} & Gain Index Policy& \multicolumn{1}{c|}{Optimal Policy} & Gain Index Policy\\ \hline
		B1                     & \multicolumn{1}{c|}{15.48}   & 15.59  & \multicolumn{1}{c|}{2.94}    & 2.94   \\ \hline
		B2                     & \multicolumn{1}{c|}{15.63}   & 15.74  & \multicolumn{1}{c|}{3.02}    & 3.04   \\ \hline
		B3                     & \multicolumn{1}{c|}{16.32}   & 16.36  & \multicolumn{1}{c|}{3.14}   & 3.16  \\ \hline
	\end{tabular}

\end{table}

Table \ref{tab:Multi-proc} compares the gain index policy with the Whittle index policy, the myopic policy, and the round-robin policy in the settings of more sources. In particular, the Whittle index policy is developed in \cite{UoI_Gongpu} for UoI-scheduling of binary Markov chains under the average cost criterion. Therefore, remote DTMCs in Table \ref{tab:Multi-proc} are all set to be binary Markov chains so that the Whittle index policy is applicable. 
The myopic policy is a greedy policy in terms of the one-step cost. That is, at the beginning of each time slot $t$, it computes for each DTMC $i$  the one-step cost  $H(X_i(t))$ and selects the  $m$ DTMCs with the largest $m$  one-step costs.
The round-robin policy, on the other hand, can be thought of as an AoI-based policy, since it is proven to be the optimal policy for minimizing the average AoI in the case of reliable channels \cite{Kadota2018TON}.
The values of $M$  and $m$  of each group are listed in the table. The results of the optimal policy are not available here because of the prohibitive computation complexity. It is quite difficult for the value iteration algorithm to converge when $M>3$. 
In Table \ref{tab:Multi-proc}, groups C1 and C2 have reliable channels, while groups C3 and C4 have unreliable channels. The results show that the gain index policy and Whittle index policy significantly outperform the other two policies in all groups. A particularly interesting observation is that the results of the gain index policy and the Whittle index policy are nearly the same in all cases. This observation motivates a question about the relationship between these two policies. However, we have no further evidence, neither theoretical nor experimental, that implies their equivalence. Analysis on their relationship awaits future study.

\begin{table}[t]
	\centering
	\caption{Average UoI of different policies under different settings}
	\label{tab:Multi-proc}
	\begin{tabular}{|c|c|c|c|c|c|}
		\hline
		Group & ($M,m$)  & Gain Index & Whittle Index & Myopic & Round-robin \\ \hline
		C1 ($\rho=1$)   & (5,2)  & 3.73         & 3.73          & 3.87          & 3.96        \\ \hline
		C2 ($\rho=1$)    & (10,3) & 7.36         & 7.37          & 7.87          & 7.73        \\ \hline
		C3 ($\rho<1$)   & (5,2)  & 3.85         & 3.85          & 4.31          & 4.25        \\ \hline
		C4 ($\rho<1$)   & (8,4)  & 5.73         & 5.73          & 6.35          & 6.52        \\ \hline
	\end{tabular}
\end{table}
\begin{table}[h]
	\centering
	\caption{Performance of the gain index policy for C-type RMABs with different cost functions.}
	\label{tab:Ctype}
	\begin{tabular}{|c|cc|cc|}
		\hline
		\multirow{2}{*}{Group} & \multicolumn{2}{c|}{The Discounted Cost Criterion}         & \multicolumn{2}{c|}{The Average Cost Criterion}          \\ \cline{2-5} 
		& \multicolumn{1}{c|}{Optimal Policy} & Gain Index & \multicolumn{1}{c|}{Optimal Policy} & Gain Index \\ \hline
		D1 ($H_1,N=3$)                     & \multicolumn{1}{c|}{13.43}   & 13.78  & \multicolumn{1}{c|}{2.68}    & 2.68   \\ \hline
		D2  ($H_1,N=4$)                  & \multicolumn{1}{c|}{20.47}   & 20.53  & \multicolumn{1}{c|}{4.07}    & 4.08   \\ \hline
		E3 ($H_2,N=3$)                 & \multicolumn{1}{c|}{5.75}   & 5.80  & \multicolumn{1}{c|}{1.13}   & 1.14 \\ \hline
		E4  ($H_2,N=4$)          & \multicolumn{1}{c|}{10.02}   & 10.08  & \multicolumn{1}{c|}{1.99}   & 2.0  \\ \hline
	\end{tabular}
\end{table}
Next, we apply the gain index policy to two C-type RMABs with different cost functions. In particular, for $X = [{x_1}, \cdots ,{x_N}] \in \Omega $, define the following cost functions:
\begin{align*}
{H_1}(X) = \sum\limits_{n = 1}^M {n{x_n}}  - \sqrt {\sum\limits_{i = 1}^M {x_i^2} },\quad {H_2}(X) = \sum\limits_{n = 1}^M {n{x_n}}  - \log \sum\limits_{n = 1}^M {{e^{{x_n}}}} .
\end{align*}
Both the above functions are concave w.r.t. $X$. As in Table \ref{tab:M2N3} and Table \ref{tab:M2N4}, we consider the setting of $M=2,m=1$  and compute the results of the optimal policy as the benchmark. The results are shown in Table \ref{tab:Ctype}. The DTMCs have 3 states in groups D1 and E1, and 4 states in groups D2 and E2. In each group, the channel is unreliable and ${\rho _1} = 0.8,{\rho _2} = 0.7$. Again, the results of the gain index policy are very close to the optimal values. The results demonstrate that the gain index policy performs well in the C-type RMABs.

\begin{table}[t]
	\centering
	\caption{Performance of the gain index policy for general RMABs.}
	\label{tab:general-RMAB}
	\begin{tabular}{|c|c|c|c|c|}
		\hline
		Group & ($M,m$) &Optimal Policy  & Gain Index & Whittle Index  \\ \hline
		F1  &  (3,1)&53.2  & 53.56         & 53.72                 \\ \hline
		F2&   (4,1)&88.7  & 91.18        & 92.06                 \\ \hline
		G1 &   (2,1)&5.03  &  5.07        &  5.08                 \\ \hline
		G2&   (3,1)&12.65  & 13.12          &12.98                   \\ \hline
	\end{tabular}
\end{table}
Finally, we show that the gain index policy can be applied to general RMABs and achieves good performance. The results are reported in Table \ref{tab:general-RMAB}. In groups F1 and F2, the RMAB is a scheduling problem that minimizes the discounted total error covariance of multiple remote Kalman filters, as studied in \cite{Shiling2020}. The RMAB in groups G1 and G2 is the minimum-AoI scheduling studied in \cite{AoIcost_nonlinear_MIT}, where the objective is to minimize the average cost of AoI. The cost functions of the two bandits in G1 are $F_1(x)=x^2$ and $F_2(x)=3^{x/4}$, where $x$ is AoI. In G2, the cost function of the third bandit is $F_3(x)=x^3/4$. The references mentioned above have developed the Whittle index policy for the RMABs. We thus present the performance of the Whittle index policy and the optimal policy as benchmarks. The results show that the gain index policy is comparable with the Whittle index policy in these applications---both policies are near-optimal.

\section{Conclusion}   \label{sec:con}
This paper adopted UoI as a metric to evaluate the information freshness of finite-state Markov sources.
We investigated the minimum-UoI scheduling problem of an information update system in which $M$  finite-state Markov sources transmit information to a remote central monitor via $m$  channels. We formulated the problem as an RMAB and studied the scheduling policies to minimize the expected discounted total UoI and the long-term average UoI. In the RMAB formulation, each bandit corresponds to a Markov source and is formulated as a belief MDP. We developed an index policy for the RMAB with the discounted cost criterion and then extended the results to the RMAB with the average cost criterion. In particular, the development of the gain index policy for both criteria follows the below steps:
\begin{itemize}
	\item [1.]	We first relaxed the RMAB and transformed the relaxed problem into a sup-min problem. Then we fixed the decision variable of the sup problem and decoupled the inner min problem into $M$  single bandit problems.
	
	\item[2.]	We analyzed the single bandit problem and obtained useful properties that allow us to solve the sup-min problem.
	
	\item[3.]	We proposed a gradient method, based on the properties established in step 2, to solve the sup-min problem and determine its optimal policy.
	
	\item[4. ]	We obtained an index policy for the original RMAB problem by rounding up the optimal policy for the sup-min problem. We also proposed an efficient algorithm to compute the index for each bandit. 
\end{itemize}
It is worth noting that the proposed gain index policy is valid not just for the UoI scheduling problem of focus here. We showed that it applies to general RMABs as long as the cost functions are bounded. Moreover, we proved that the gain index policy is asymptotically optimal as $m$  and $M$ tend to $\infty$  with $m/M$  fixed. Numerical results also demonstrated the excellent performance of our policy. The universality and asymptotic optimality of the gain index policy make it an excellent method for RMABs.
	
In this paper, we studied the UoI scheduling problem in a general setting of finite-state Markov sources. Further generalization of the UoI metric to Markov sources with continuous state space will make this metric applicable to a broader range of applications. A natural way to do this is to define UoI using differential entropy. This is an interesting problem for future research.

\section*{Appendix I}  \label{Apd:A}
\subsection{Proof of Lemma \ref{Lem:Vxconcave}}
This part presents the proof of Lemma \ref{Lem:Vxconcave}, which is stated in Section \ref{subsec:disc-single}.
\begin{IEEEproof}
	We prove the concavity of $V(X)$  by the value iteration algorithm and mathematic induction. Let ${V^0}(X) = 0$ for all $X\in \Omega$. For $n\ge 1$ and $X\in \Omega$, define the following iteration:
	\begin{align} \label{eq:Lem1-VI}
	{V^{n + 1}}(X) = \min \left\{ {H(X) + \lambda  + \beta \rho \mathop \sum \limits_{i = 1}^N {x_i}{V^n}({{\bf{T}}_i}) + \beta (1 - \rho ){V^n}({\bf{T}}X),H(X) + \beta {V^n}({\bf{T}}X)} \right\}.
	\end{align}
	Let
	\begin{align*}
	V_0^n(X) &= H(X) + \beta {V^n}({\bf{T}}X),\\
	V_1^n(X) &= H(X) + \lambda  + \beta \rho \mathop \sum \limits_{i = 1}^N {x_i}{V^n}({{\bf{T}}_i}) + \beta (1 - \rho ){V^n}({\bf{T}}X ).
	\end{align*}
	Assume that ${V^n}(X)$ is concave, then ${V^n}({\bf{T}}X)$ is concave w.r.t. $X$. Since $H(X)$ is also a concave function, it is easy to verify that $V_0^n(X)$ and $V_1^n(X)$ are concave. We then have
	\begin{align}
	{V^{n + 1}}(X) = \min \left\{ {V_1^n(X),V_0^n(X)} \right\},\quad X \in \Omega .
	\end{align}
	The point-wise minimum of two concave functions is still concave, hence $V^{n+1}(X)$  is concave.
	Since $V^0(X)$  is concave, using the mathematic induction argument yields that $V^n(X)$  generated by \eqref{eq:Lem1-VI} is concave for all $n$. It is well-known that the value iteration converges to the optimal value function \cite{puterman1994markov}, i.e., $\mathop {\lim }\limits_{n \to \infty } {V^n}(X) = V(X)$; we thus conclude that the optimal value function $V(X)$  is concave.
\end{IEEEproof}

\subsection{Proof of Lemma 2}
This part presents the proof of Lemma \ref{Lem:ConvexAct} stated in Section \ref{subsec:disc-single}.

\begin{IEEEproof}
	Define a function
	\begin{align}
	d(X) = r(X) - a(X) = \beta \rho V({\bf{T}}X) - \lambda  - \beta \rho \mathop \sum \limits_{i = 1}^N {x_i}V({{\bf{T}}_i}).
	\end{align}
	It is well-known that composition with an affine mapping preserves concavity. Therefore, $V(X)$ is concave implies that $V({\bf{T}}X)$  is also concave.  Then it is easy to see that  $d(X)$ is a concave function of $X$. The optimal policy for the single bandit problem is to take active action in the set $\mathcal{A}_\lambda=\{X:d(X)\ge 0, X\in \Omega \}$.
	
	 The concavity of  $d(X)$ implies that  $\mathcal{A}_\lambda$  is convex. To see this, let $\mathcal{A}^b_\lambda=\{X:d(X)= 0, X\in \Omega \}$. Since $d(X)$ is concave, for any ${X_1}, \cdots ,{X_k} \in \mathcal{A}_\lambda ^b$ and ${\alpha _1}, \cdots ,{\alpha _k} \in [0,1]$ such that $\sum\nolimits_{i = 1}^k {{\alpha _i}}  = 1$, we have
	\begin{align} \label{eq:Lem2-concave}
	0 = \sum\limits_{i = 1}^k {{\alpha _i}d({X_i})}  \le d\left( {\sum\limits_{i = 1}^k {{\alpha _i}{X_i}} } \right).
	\end{align}
	Denote by $Conv{\rm{ }}\mathcal{A}_\lambda ^b = \left\{ {\sum\nolimits_{i = 1}^k {{\alpha _i}} {X_i}|{X_i} \in \mathcal{A}_\lambda ^b,{\alpha _i} \in [0,1],\sum\nolimits_{i = 1}^k {{\alpha _i}}  = 1} \right\}$ the convex hull of $\mathcal{A}^b_\lambda$. Then \eqref{eq:Lem2-concave} means that  $Conv{\rm{ }}\mathcal{A}_\lambda ^b \subseteq \mathcal{A}_\lambda$.
	
	Further, for any  $Y\notin Conv{\rm{ }}\mathcal{A}_\lambda ^b$, we can find a point $X\in \mathcal{A}^b_\lambda$, a point $Z\in Conv{\rm{ }}\mathcal{A}_\lambda ^b$, and $\alpha\in (0,1)$ such that $X = \alpha Z + (1 - \alpha )Y$. Then
	\begin{align}
	0=d(X) = d\left( {\alpha Z + (1 - \alpha )Y} \right) \ge \alpha d(Z) + (1 - \alpha )d(Y).
	\end{align}
	Since $d(Z)\ge 0$, the above inequality implies $d(Y)\le 0$. On the other hand, $d(Y)\neq 0$ because $Y\notin \mathcal{A}^b_\lambda \subseteq Conv{\rm{ }}\mathcal{A}_\lambda ^b$. Then we must have $d(Y)<0$. This means that, for any $Y\notin Conv{\rm{ }}\mathcal{A}_\lambda ^b$, we have $Y\notin \mathcal{A}_\lambda$. That is,  $\mathcal{A}_\lambda \subseteq Conv{\rm{ }}\mathcal{A}_\lambda ^b$. 
	
	Putting the above results together yields $\mathcal{A}_\lambda = Conv{\rm{ }}\mathcal{A}_\lambda ^b$. That is,  $\mathcal{A}_\lambda$  is a convex set.
	
\end{IEEEproof}

\subsection{Proof of Corollary 1}
This part presents the proof of Corollary \ref{Coro:lambda0}, which is stated in Section \ref{subsec:disc-single}.
\begin{IEEEproof}
	Let ${e_i} = {[0 \cdots 1 \cdots 0]^T}\in \mathbb{R}^N$ denote the vector whose  $i$-th element is 1 and other elements are 0. Note that if $\lambda=0$ then $a({e_i}) = \beta V({{\bf{T}}_i}) = r({e_i})$ for  all $i\in [N]$, implying that the belief states  $\{ e_i:i \in [N]\} $ are on the boundary of the optimal active set. According to Lemma \ref{Lem:ConvexAct}, the optimal active set is always a convex set. While the convex hull of  $\{ e_i:i \in [N]\} $ is $\Omega$. Hence the whole belief state space belongs to the active set.
	
	For statement 2, note that the entropy function  $H(X)$ is bounded in $\Omega$. Then the policy that takes passive action in all belief states, denoted by $o$, has the following value function:
	\begin{align}
	{V_o}(X,\lambda ) = \sum\limits_{t = 1}^\infty  {{\beta ^{t - 1}}H({{\bf{T}}^{t - 1}}X)}  < \infty.
	\end{align}
	For an arbitrary nonempty set $\kappa\subseteq \Omega$, we also refer to  $\kappa$ as the policy that takes active action for $X\in \kappa$  and passive action for $X\notin \kappa$. Using a similar argument as in \eqref{eq:lem3-1}, the value function of policy $\kappa$ is the expected discounted total reward obtained by this policy. That is,
	\begin{align}
	{V_\kappa }(X,\lambda ) = {E_\kappa }\left[ {\sum\limits_{t = 1}^\infty  {{\beta ^{t - 1}}H\left( {X(t)} \right)|X} } \right] + \lambda {E_\kappa }\left[ {\sum\limits_{t = 1}^\infty  {{\beta ^{t - 1}}u(t)|X} } \right],
	\end{align}
	where ${E_\kappa }[ \cdot |X]$ is the expectation taken over the Markov chain generated by policy  $\kappa$ with initial state $X$. It follows immediately that
	\begin{align} \label{eq:Cor-phiV}
	\frac{{\partial {V_\kappa }(X,\lambda )}}{{\partial \lambda }} = {E_\kappa }\left[ {\sum\limits_{t = 1}^\infty  {{\beta ^{t - 1}}u(t)|X} } \right] \ge 0.
	\end{align}
	For any $X\in \Omega$, if the inequality in \eqref{eq:Cor-phiV} is satisfied with equality, then it means that, starting from belief state $X$, policy $\kappa$  will take passive action in all the following time; hence
	\begin{align}
	{V_\kappa }(X,\lambda ) = \sum\limits_{t = 1}^\infty  {{\beta ^{t - 1}}H({{\bf{T}}^{t - 1}}X)}  = {V_o}(X,\lambda ).
	\end{align}
	On the other hand, if $\partial {V_\kappa }(X,\lambda )/\partial \lambda $ strictly positive for some $X$, then ${V_\kappa }(X,\lambda )$  is an increasing function of $\lambda$; hence there must exist a finite  $\lambda_X$ such that
	\[{V_\kappa }(X,{\lambda _X}) > {V_o}(X,{\lambda _X}).\]
	The above inequality also holds for $\lambda > \lambda_X$. Taking ${\lambda ^\kappa } = \max \{ {\lambda _X}:\partial {V_\kappa }(X,\lambda )/\partial \lambda  > 0,X \in \Omega \} $, then policy $o$ is better than $\kappa$ for any $\lambda > \lambda^\kappa$. It follows that policy $o$  is the optimal policy when $\lambda  > \bar \lambda  = {\max _\kappa }\left\{ {{\lambda ^\kappa }} \right\}$.
\end{IEEEproof}

\section*{Appendix II}  \label{Apd:B}
\subsection*{A. Proof of Theorem 2}
As stated in the sketch of the proof under Theorem 2, the desired result is proved based on an auxiliary MDP $J'(\lambda)$. For simplicity, we do not repeat the construction of $J'(\lambda)$ here. The full proof below will proceed based on $J'(\lambda)$ constructed in the proof sketch.
\begin{IEEEproof}
	Consider an arbitrary $\lambda\in [0,\infty)$. Since $J'(\lambda)$ and $J(\lambda)$ have the same state space and action space, a policy of $J(\lambda)$  can also be applied to $J'(\lambda)$. Let $\pi$ denote a policy of $J(\lambda)$ and $\pi(X)\in \{0,1\}$ denote the action taken in state $X$ by policy $\pi$. Define
	\begin{align*}
	{C_\pi }(X) = H(X) + \lambda \pi (X),\quad X \in {\cal S},
	\end{align*}
	and
	\begin{align*}
	C_\pi ^L(X) =  \begin{cases}
	H(X) + \lambda \pi (X),\quad &X \in {{\cal S}^L}\\
	H(\omega ) + \lambda \pi (X),\quad &X \notin {{\cal S}^L}.
	\end{cases} 
	\end{align*}
	Further, let $\mathbf{P}_\pi$  and $\mathbf{Q}_\pi$  denote the transition matrices of $J(\lambda)$  and $J'(\lambda)$  under policy $\pi$, respectively. Then the value function of  $J(\lambda)$ under policy $\pi$  is determined by
	\begin{align} \label{eq:Thm2-Vpi}
	{V_\pi }(X) = {C_\pi }(X) + \beta \sum\limits_{Y \in {\cal S}} {{\mathbf{P}_\pi }(X,Y){V_\pi }(Y)} ,\quad X \in {\cal S}.
	\end{align}
	The above equation can be written in vector form as
	\begin{align} \label{eq:thm2-P}
	{V_\pi } = {C_\pi } + \beta {\mathbf{P}_\pi }{V_\pi }.
	\end{align}
	Similarly, the value function of  $J'(\lambda)$ under policy $\pi$  is given by
	\begin{align} \label{eq:thm2-Q}
	{\varphi _\pi } = C_\pi ^L + \beta {\mathbf{Q}_\pi }{\varphi _\pi }.
	\end{align}
	From \eqref{eq:thm2-P} and \eqref{eq:thm2-Q},
	\begin{align}
	{V_\pi } - {\varphi _\pi } 
	= {C_\pi } - C_\pi ^L + \beta \left( {{\mathbf{P}_\pi } - {\mathbf{Q}_\pi }} \right){V_\pi } + \beta {\mathbf{Q}_\pi }\left( {{V_\pi } - {\varphi _\pi }} \right).
	\end{align}
	Since $\mathbf{Q}_\pi$ is a stochastic matrix and $\beta\in [0,1)$, $\left( {\mathbf{I} - \beta {\mathbf{Q}_\pi }} \right)$ is invertible. It follows that
	\begin{align} \label{eq:thm2-Vphi}
	{V_\pi } - {\varphi _\pi } = {\left( {\mathbf{I} - \beta {\mathbf{Q}_\pi }} \right)^{ - 1}}\left( {{C_\pi } - C_\pi ^L} \right) + \beta {\left( {\mathbf{I} - \beta {\mathbf{Q}_\pi }} \right)^{ - 1}}\left( {{\mathbf{P}_\pi } - {\mathbf{Q}_\pi }} \right){V_\pi }.
	\end{align}
	By definition, ${C_\pi }(X) - C_\pi ^L(X) = 0$ if $X\in \mathcal{S}$; otherwise, $X = {\bf{T}}_i^{L + k}$ for some $i\in [N]$ and $k\ge 1$, and
	\begin{align} \label{eq:thm2-Vphi2}
	\left| {{C_\pi }(X) - C_\pi ^L(X)} \right| = \left| {H({\bf{T}}_i^{L + k}) - H(\omega )} \right| \le {\sigma _L}.
	\end{align}
	Therefore,
	\begin{align}
	\left|\left| {{{\left( {\mathbf{I} - \beta {\mathbf{Q}_\pi }} \right)}^{ - 1}}\left( {{C_\pi } - C_\pi ^L} \right)} \right|\right| = \left|\left| {{C_\pi } - C_\pi ^L + \sum\limits_{n = 1}^\infty  {{\beta ^n}\mathbf{Q}_\pi ^n} \left( {{C_\pi } - C_\pi ^L} \right)} \right|\right| \le \left|\left| {{C_\pi } - C_\pi ^L} \right|\right| + \frac{{\beta {\sigma _L}}}{{1 - \beta }}{e},
	\end{align}
	where  $e$ is the all-one vector. For the second term of \eqref{eq:thm2-Vphi}, first note that $V_\pi$ is bounded. In particular, for $X\in \mathcal{S}$,
	\begin{align}
	{V_\pi }(X) = \sum\limits_{n = 0}^\infty  {{\beta ^n}\sum\limits_{Y \in {\cal S}} {\mathbf{P}_\pi ^n(X,Y){C_\pi }(Y)} }  \le \sum\limits_{n = 0}^\infty  {{\beta ^n}\sum\limits_{Y \in {\cal S}} {\mathbf{P}_\pi ^n(X,Y)\left[ {{B_H} + \lambda } \right]} }  = \frac{{{B_H} + \lambda }}{{1 - \beta }}.
	\end{align}
	Meanwhile, $\left| {{\mathbf{P}_\pi }(X,Y) - {\mathbf{Q}_\pi }(X,Y)} \right| \le \rho {\eta _L}$ for all $X,Y\in \mathcal{S}$, and each column of  ${{\mathbf{P}_\pi } - {\mathbf{Q}_\pi }}$ has at most $N$  non-zero elements. Let $b_\pi = ({{\mathbf{P}_\pi } - {\mathbf{Q}_\pi }})V_\pi$ , then we have
	\begin{align} \label{eq:thm2-bpi}
	\left| {{b_\pi }(X)} \right| \le N\rho {\eta _L}\frac{{{B_H} + \lambda }}{{1 - \beta }},\quad X \in {\cal S}.
	\end{align}
	From \eqref{eq:thm2-Vphi}-\eqref{eq:thm2-bpi}, for any $X\in \mathcal{S}^L$,
	\begin{align} \notag
	\left| {{V_\pi }(X) - {\varphi _\pi }(X)} \right| &\le \left| {C_\pi }(X) - C_\pi ^L(X)\right| + \frac{{\beta {\sigma _L}}}{{1 - \beta }} + \left|\beta \sum\limits_{n = 0}^\infty  {{\beta ^n}\sum\limits_{Y \in {\cal S}} {\mathbf{Q}_\pi ^n(X,Y){b_\pi }(Y)} }  \right|\\
	&\le \frac{{\beta {\sigma _L}}}{{1 - \beta }} + N\rho {\eta _L}\frac{{{B_H} + \lambda }}{{1 - \beta }}\beta \sum\limits_{n = 0}^\infty  {{\beta ^n}}  = \frac{{\beta {\sigma _L}}}{{1 - \beta }} + \beta \rho {\eta _L}N\frac{{{B_H} + \lambda }}{{{{\left( {1 - \beta } \right)}^2}}}.
	\end{align}
	Finally, suppose that $\pi_1$ and $\pi_2$ are optimal policies of $J(\lambda)$ and $J'(\lambda)$, respectively. For any $X\in \mathcal{S}^L$,
	\begin{align} \label{eq:thm2-38}
	{V_{{\pi _2}}}(X) \ge {V_{{\pi _1}}}(X) \ge {\varphi _{{\pi _1}}}(X) - \frac{{\beta {\sigma _L}}}{{1 - \beta }} - \beta \rho {\eta _L}N\frac{{{B_H} + \lambda }}{{{{\left( {1 - \beta } \right)}^2}}} \ge {\varphi _{{\pi _2}}}(X) - \frac{{\beta {\sigma _L}}}{{1 - \beta }} - \beta \rho {\eta _L}N\frac{{{B_H} + \lambda }}{{{{\left( {1 - \beta } \right)}^2}}}.
	\end{align}
	Applying a similar argument yields
	\begin{align} \label{eq:thm2-39}
	{\varphi _{{\pi _2}}}(X) \ge {V_{{\pi _1}}}(X) - \frac{{\beta {\sigma _L}}}{{1 - \beta }} - \beta \rho {\eta _L}N\frac{{{B_H} + \lambda }}{{{{\left( {1 - \beta } \right)}^2}}}.
	\end{align}
	Then \eqref{eq:thm2-38} and \eqref{eq:thm2-39} imply that
	\begin{align}
	\left| {{V_{{\pi _1}}}(X) - {\varphi _{{\pi _2}}}(X)} \right| = \left| {V(X,\lambda ) - \varphi (X,\lambda )} \right| \le \frac{{\beta {\sigma _L}}}{{1 - \beta }} + \beta \rho {\eta _L}N\frac{{{B_H} + \lambda }}{{{{\left( {1 - \beta } \right)}^2}}}.
	\end{align}
	As discussed in the proof sketch, $\varphi (X) = {\phi ^L}(X)$ for $X\in \mathcal{S}^L$. Therefore $\left| {V(X) - {\phi ^L}(X)} \right|$ also satisfies the above inequality.
	
\end{IEEEproof}

\section*{Appendix III}
\subsection*{A. Proof of Lemma \ref{Lem:g-pro}}
\begin{IEEEproof}
	For any  $\pi \in \Gamma$, $g_\pi(\lambda)$ and the differential value function $Z_\pi(X,\lambda)$ can be evaluated by
	\begin{align} \label{eq:PE-ave}
	{Z_\pi }(X,\lambda ) + {g_\pi }(\lambda ) = H(X) + \lambda \pi (X) + \sum\limits_{Y \in {\cal S}} {{P_\pi }(Y|X){Z_\pi }(Y,\lambda )} ,\quad X \in {\cal S}.
	\end{align}
	Taking derivatives w.r.t. $\lambda$  for both sides yields
	\begin{align} \label{eq:lem5-dZ}
	\frac{{\partial {Z_\pi }(X,\lambda )}}{{\partial \lambda }} + {g'_\pi }(\lambda ) = \pi (X) + \sum\limits_{Y \in {\cal S}} {{P_\pi }(Y|X)\frac{{\partial {Z_\pi }(X,\lambda )}}{{\partial \lambda }}} ,\quad X \in {\cal S}.
	\end{align}
	Policy $\pi$ can also be applied to the auxiliary MDP (see Definition \ref{def:auxMDP}) of this single bandit problem. The corresponding policy evaluation equations are given by
	\begin{align} \label{eq:lem5-z}
	{z_\pi }(X) + g_\pi ^a = \pi (X) + \sum\limits_{Y \in {\cal S}} {{P_\pi }(Y|X){z_\pi }(Y)} ,\quad X \in {\cal S}.
	\end{align}
	Note that \eqref{eq:lem5-dZ} and \eqref{eq:lem5-z} are of the same form. It is known that the set of equations has a unique solution up to adding a constant to $z_\pi(\cdot)$. Hence $g_\pi ^a = {g'_\pi }(\lambda )$ for all $\lambda \ge 0$. We set $z({{\bf{T}}_1} ) = 0$  because of the assumption of $Z({{\bf{T}}_1},\lambda ) = 0$ implies that $\partial Z({{\bf{T}}_1},\lambda )/\partial \lambda  = 0$. In fact, we can select an arbitrary  $X\in \mathcal{S}$ and let  $z(X)=0$ as far as ${g'_\pi }$  is concerned. This proves statement 1. 
	
	Because $g(\lambda)$  is the minimum average cost. Then for any $\lambda \ge 0$,
	\begin{align} \label{eq:lem5-g}
	g(\lambda ) = \mathop {\min }\limits_{\pi  \in \Gamma } {g_\pi }(\lambda ).
	\end{align}
	Statement 1 implies that $g_\pi(\lambda)$  is a linear function for any $\pi  \in \Gamma$. It then follows from \eqref{eq:lem5-g} that $g(\lambda)$  is piecewise linear and concave.
	To show that $g(\lambda)$  is increasing, we invoke the property that ${g'_\pi }(\lambda )=g_\pi ^a $ for any $\pi \in \Gamma$. Note that $g^a_\pi$ determined by \eqref{eq:lem5-z} is the average cost of the auxiliary MDP and that the cost function of the auxiliary MDP is non-negative. Hence $g^a_\pi\ge 0$ for all $\pi \in \Gamma$. As a result, $g'(\lambda) = g'_{\pi^*}(\lambda)\ge 0$, where $\pi^*$ is an optimal policy in $\Gamma$. Statement 2 follows immediately.  
	
	Statement 3 can also be verified by \eqref{eq:lem5-g}. If $G(\lambda)$ has a unique optimal policy for a given $\lambda$, say $\pi_\lambda$, then $g'(\lambda ) = {g'_{{\pi _\lambda }}}(\lambda )$ always exists. If $G(\lambda)$ has $k\ge 2$ optimal policies for $\lambda=y$, say ${\Pi _y} = \{ {\pi _i}:i \in [k]\} $, it means that the $k$ linear functions $\{ {g_{{\pi _i}}}(\lambda ),i \in [k]\} $ intersect at the point $y$. Without loss of generality, let us assume that ${g'_{{\pi _i}}}(\lambda ) \ge {g'_{{\pi _{i + 1}}}}(\lambda )$ for all $i\in [k-1]$. Then, by the piecewise linearity of $g(\lambda)$, there exists a $\sigma > 0$ such that 
	\begin{align*}
	g(\lambda ) = \mathop {\min }\limits_{\pi  \in \Gamma } {g_\pi }(\lambda ) = \mathop {\min }\limits_{i \in [k]} {g_{{\pi _i}}}(\lambda ) =  \begin{cases}
	{g_{{\pi _1}}}(\lambda ),{\text{ if }}\lambda  \in (y - \sigma ,y)\\
	{g_{{\pi _k}}}(\lambda ),{\text{ if }}\lambda  \in (y,y + \sigma ).
	\end{cases} 
	\end{align*}
	Hence $g'({y_ - }) = {g'_{{\pi _1}}}({y_ - })$ and $g'({y_ + }) = {g'_{{\pi _k}}}({y_ + })$. The desired result follows immediately.
	
\end{IEEEproof}

\subsection*{B. Proof of Lemma \ref{Lem:g-coro}}

This part presents the proof of Lemma \ref{Lem:g-coro}, which is stated in Section \ref{subsec:ave-single}.
\begin{IEEEproof}
	We use another conclusion in Dutta’s theorem \cite{dutta1991discounted} to prove this lemma. In particular, let ${\pi _\lambda }( \cdot ,\beta )$ denote an optimal policy for the discounted cost problem $J(\lambda)$ with discount factor $\beta\in [0,1)$. For $X\in \mathcal{S}$, ${\pi _\lambda }(X,\beta ) \in \{ 0,1\} $ denotes the action taken by this policy in state $X$. From Corollary 1 and its proof, it is easy to see that ${V^\beta }(X,\lambda )$ is bounded for all $\beta\in [0,1)$ and $\lambda < \infty$. Then Dutta's theorem says that
	
	\textit{If a series of optimal policies ${\pi _\lambda }( \cdot ,\beta )$  for the belief MDP $J(\lambda)$  with discount factor $\beta_k$  pointwise converges to a limit $\pi_\lambda(\cdot)$ as $\beta_k \to 1$, then $\pi_\lambda(\cdot)$ is an optimal policy for the average cost problem $G(\lambda)$.
	}

	We have proved in Corollary \ref{Coro:lambda0} that $\pi_0(X,\beta)=1$ for all $X\in \mathcal{S}$ and $\beta \in [0,1)$. That is,
	\begin{align*}
	\mathop {\lim }\limits_{\beta  \uparrow 1} {\pi _0}(X,\beta ) = {\pi _0}(X) \buildrel \Delta \over = 1,\quad \forall X \in {\cal S}.
	\end{align*}
	Therefore, $\pi_0(\cdot)$ is an optimal policy for $G(0)$. According to statement 1 of Lemma \ref{Lem:g-pro}, $g'(\lambda ) = {g'_{{\pi _0}}}(\lambda )$ is the average cost of the auxiliary MDP under policy $\pi_0(\cdot)$.  Since $\pi_0(\cdot)$ is  the policy that takes active action in all belief states, it is clear that the average cost of the auxiliary MDP is 1. This proves statement 1. Statement 2 can be proved similarly. 

\end{IEEEproof}

\subsection*{C. Proof of Theorem \ref{Thm:bound-ave}}
This part presents the proof of Theorem  \ref{Thm:bound-ave}, which is stated in Section \ref{subsec:ave-single}.
\begin{IEEEproof}
	We establish the desired bound based on the construction and some intermediate results in the proof of Theorem \ref{thm:errorbound1}. For any $\pi \in \Gamma$, let $V^\beta_\pi(X)$ denote the value function of $J(\lambda)$ with discount factor $\beta$ under the policy $\pi$ (i.e., $V_\pi(X)$ in \eqref{eq:Thm2-Vpi}, we use the notation $V^\beta_\pi(X)$ here to emphasize the value of $\beta$). Likewise, let $\varphi _\pi ^\beta (X)$ denote $\varphi_\pi (X)$ with discount factor $\beta$ in \eqref{eq:thm2-Q}. It is well-known that (see, e.g., \cite{puterman1994markov})
	\begin{align}
	{g_\pi }(\lambda ) = {g_\pi }(X,\lambda ) = \mathop {\lim }\limits_{\beta  \uparrow 1} (1 - \beta )V_\pi ^\beta (X,\lambda ),\quad \forall X \in {\cal S}.
	\end{align}
	Similarly, 
	\begin{align}
	g_\pi ^L(\lambda ) = g_\pi ^L(X,\lambda ) = \mathop {\lim }\limits_{\beta  \uparrow 1} (1 - \beta )\varphi _\pi ^\beta (X,\lambda ),\quad \forall X \in {{\cal S}^L}.
	\end{align}
	Recall that ${{\cal S}^L}$ is the belief state space of $G^L(\lambda)$. It follows that, for any $X\in \mathcal{S}^L$,
	\begin{align*}
	{g_\pi }(\lambda ) - g_\pi ^L(\lambda ) = \mathop {\lim }\limits_{\beta  \uparrow 1} (1 - \beta )\left[ {V_\pi ^\beta (X,\lambda ) - \varphi _\pi ^\beta (X,\lambda )} \right].
	\end{align*}
	Let $V_\pi ^\beta (\lambda )$  and $\varphi _\pi ^\beta (\lambda )$   denote respectively the vector form of  $V_\pi ^\beta (X,\lambda )$ and $\varphi _\pi ^\beta (X,\lambda )$  over $X \in {{\cal S}^L}$. Then according to \eqref{eq:thm2-Vphi},
	\begin{align} \notag 
	\mathop {\lim }\limits_{\beta  \uparrow 1} (1 - \beta )\left[ {V_\pi ^\beta ( \lambda ) - \varphi _\pi ^\beta ( \lambda )} \right] &= \mathop {\lim }\limits_{\beta  \uparrow 1} (1 - \beta ){\left( {\mathbf{I} - \beta {\mathbf{Q}_\pi }} \right)^{ - 1}}\left( {{C_\pi } - C_\pi ^L} \right)\\ \label{eq:thm4-lim}
	&+ \mathop {\lim }\limits_{\beta  \uparrow 1} (1 - \beta )\beta {\left( {\mathbf{I} - \beta {\mathbf{Q}_\pi }} \right)^{ - 1}}\left( {{\mathbf{P}_\pi } - {\mathbf{Q}_\pi }} \right)V_\pi ^\beta (\lambda ).
	\end{align}
	Let $\xi$ be the equilibrium distribution of the Markov chain associated with $\mathbf{Q}_\pi$. Then we have $\xi^\top  = \xi^\top \mathbf{Q}_\pi$ and
	\begin{align}
	{g_\pi }(\lambda ) - g_\pi ^L(\lambda ) = [{g_\pi }(\lambda ) - g_\pi ^L(\lambda )]\xi^\top e = \mathop {\lim }\limits_{\beta  \uparrow 1} (1 - \beta )\xi^\top \left[ {V_\pi ^\beta (X,\lambda ) - \varphi _\pi ^\beta (X,\lambda )} \right].
	\end{align}
	where $e$  is the all-one vector. Note that
	\begin{align}\notag
	\mathop {\lim }\limits_{\beta  \uparrow 1} (1 - \beta )\beta \xi^\top{\left( {\mathbf{I} - \beta {\mathbf{Q}_\pi }} \right)^{ - 1}}\left( {{\mathbf{P}_\pi } - {\mathbf{Q}_\pi }} \right)V_\pi ^\beta ( \lambda ) &= \mathop {\lim }\limits_{\beta  \uparrow 1} (1 - \beta )\beta  \sum_{t=0}^{\infty} \beta^t\xi^\top \mathbf{Q}^t_\pi \left( {{\mathbf{P}_\pi } - {\mathbf{Q}_\pi }} \right)V_\pi ^\beta ( \lambda ) \\ \notag
	&= \sum_{t=0}^{\infty} \xi^\top\left( {{\mathbf{P}_\pi } - {\mathbf{Q}_\pi }} \right)\mathop {\lim }\limits_{\beta  \uparrow 1} (1 - \beta )V_\pi ^\beta (\lambda )\\ \notag
	&= \sum_{t=0}^{\infty} \xi^\top\left( {{\mathbf{P}_\pi } - {\mathbf{Q}_\pi }} \right){g_\pi }(\lambda ){e}\\ 
	&= {g_\pi }(\lambda )\sum_{t=0}^{\infty} \xi^\top\left( {{\mathbf{P}_\pi }{e} - {\mathbf{Q}_\pi }{e}} \right) = 0.
	\end{align}
	The last equality holds because $\mathbf{P}_\pi$  and $\mathbf{Q}_\pi$ are stochastic matrices. For the first term in \eqref{eq:thm4-lim}, 
	\begin{align} \notag
	\left|{\mathop {\lim }\limits_{\beta  \uparrow 1} (1 - \beta )\xi^\top{{\left( {\mathbf{I} - \beta {\mathbf{Q}_\pi }} \right)}^{ - 1}}\left( {{C_\pi } - C_\pi ^L} \right)} \right|&\le
	\left|\left| {\mathop {\lim }\limits_{\beta  \uparrow 1} (1 - \beta ){{\left( {\mathbf{I} - \beta {\mathbf{Q}_\pi }} \right)}^{ - 1}}\left( {{C_\pi } - C_\pi ^L} \right)} \right|\right|_\infty \\ \notag
	&= \mathop {\lim }\limits_{\beta  \uparrow 1} (1 - \beta )\left|\left| {\sum\limits_{n = 0}^\infty  {{\beta ^n}\mathbf{Q}_\pi ^n} \left( {{C_\pi } - C_\pi ^L} \right)} \right|\right|_\infty\\ \label{eq:thm4-bound}
	&\le \mathop {\lim }\limits_{\beta  \uparrow 1} (1 - \beta )\left[ {\left|\left| {{C_\pi } - C_\pi ^L} \right|\right|_\infty + \frac{{\beta {\sigma _L}}}{{1 - \beta }}} \right]
	= {\sigma _L}.
	\end{align}
	Putting together \eqref{eq:thm4-lim}-\eqref{eq:thm4-bound} yields the desired result.
\end{IEEEproof}

\section*{Appendix IV}
\subsection*{A. Proof of Lemma \ref{lem:7}}
\begin{IEEEproof}
	Note that the OR policy is stationary and that all bandits are mutually independent under this policy; hence each bandit of the RMAB governed by the OR policy reduces to a Markov chain. Conditioning on the assumption of the initial state following the equilibrium distribution (Assumption 1), the state distribution of the RMAB under the OR policy is at equilibrium at any time. Hence the probability of the OR policy taking active action for a particular bandit is independent of time. That is, for any $i\in [M]$,
	\begin{align*}
	\Pr \{ {u_i}(t) = 1\}  = \Pr \left\{ {{a_i}\left( {{X_i}(t),{\lambda ^*}} \right) \le {r_i}\left( {{X_i}(t),{\lambda ^*}} \right)} \right\},
	\end{align*}
	and the distribution of $X_i(t)$ is independent of time. By definition, bandits of the same class are stochastically identical, hence $\Pr \{ {u_i}(t) = 1\}  = \Pr \{ {u_j}(t) = 1\} $ if $i,j\in\mathcal{Q}_k$ for any $k\in [Q]$ and $t$. We suppose that $\Pr \{ {u_i}(t) = 1\}  = \alpha_k$ for $i\in \mathcal{Q}_k$. Clearly, $\alpha_k\in [0,1]$.
	
	Using \eqref{eq:pi*} and the fact that $u_i(t)\in \{0,1\}$, we have
	\begin{align*}
	\frac{\alpha }{{1 - \beta }} &= \frac{1}{M}{E_{{\pi ^*}}}\left[ {\sum\limits_{i = 1}^M {\sum\limits_{t = 1}^\infty  {{\beta ^{t - 1}}{u_i}(t)} } } \right] = \frac{1}{M}\sum\limits_{i = 1}^M {\sum\limits_{t = 1}^\infty  {{\beta ^{t - 1}}\Pr \{ {u_i}(t) = 1\} } } \\
	&= \frac{1}{M}\sum\limits_{k = 1}^Q {\sum\limits_{i \in {{\cal Q}_k}} {\sum\limits_{t = 1}^\infty  {{\beta ^{t - 1}}\Pr \{ {u_i}(t) = 1\} } } } \\
	&= \frac{1}{M}\sum\limits_{k = 1}^Q {\sum\limits_{i \in {{\cal Q}_k}} {\frac{{{\alpha _k}}}{{1 - \beta }}} } \\
	&= \sum\limits_{k = 1}^Q {\frac{{M{q_k}}}{M}\frac{{{\alpha _k}}}{{1 - \beta }}} .
	\end{align*}
	It follows that $\sum_{k=1}^Q q_k \alpha_k =\alpha\ $.
	
\end{IEEEproof}

\subsection*{B. Proof of Proposition 1}
\begin{IEEEproof}
	Note that
	\begin{align*}
	\frac{{y_t^{(M)}}}{M} = \frac{1}{M}\sum\limits_{k = 1}^Q {\sum\limits_{i \in {{\cal Q}_k}} {{u_i}(t)} }  = \sum\limits_{k = 1}^Q {{q_k}\frac{1}{{M{q_k}}}\sum\limits_{i \in {{\cal Q}_k}} {{u_i}(t)} } .
	\end{align*}
	As stated in Lemma \ref{lem:7}, $E[{u_i}(t)] = \Pr \{ {u_i}(t) = 1\}  = {\alpha _k}$ for any $i\in \mathcal{Q}_k$. Then the variance of $u_i(t)$ is $Var[u_i(t)] = (1-\alpha_k)\alpha_k$ for any $i\in \mathcal{Q}_k$. According to the central limit theorem, as $M\to \infty$,
	\begin{align*}
	\frac{1}{{M{q_k}}}\sum\limits_{i \in {{\cal Q}_k}} {{u_i}(t)} \overset{a.s.}{\longrightarrow} {\cal N}\left( {{\alpha _k},\frac{{{\alpha _k}(1 - {\alpha _k})}}{{M{q_k}}}} \right).
	\end{align*}
	It follows immediately that ${y_t^{(M)}/}{M}$ converges almost surely to a normal distribution with mean $\sum\nolimits_{k = 1}^Q {{q_k}{\alpha _k} = \alpha } $ and variance
	\begin{align*}
	{\bar \sigma ^2} = \sum\limits_{k = 1}^Q {q_k^2\frac{{{\alpha _k}(1 - {\alpha _k})}}{{M{q_k}}}}  \le \sum\limits_{k = 1}^Q {{q_k}\frac{1}{{4M}}}  = \frac{1}{{4M}}.
	\end{align*}
	The above inequality holds because $\alpha_k\in [0,1]$. Therefore, for any positive integer $k$,
	\begin{align*}
	\Pr \left\{ {\left| {\frac{{y_t^{(M)}}}{M} - \alpha } \right| \le \frac{k}{{\sqrt {4M} }}} \right\} \ge \Pr \left\{ {\left| {\frac{{y_t^{(M)}}}{M} - \alpha } \right| \le k\bar \sigma } \right\} = \Phi (k) - \Phi ( - k)\quad {\text{as }}M \to \infty.
	\end{align*}
	This completes the proof.
	
\end{IEEEproof}

\subsection*{C. Proof of Theorem \ref{thm:optimality}}
This part proves the asymptotic optimality of the gain index policy for the discounted cost problem, as stated in Theorem \ref{thm:optimality}.
\begin{IEEEproof}
	Let $\hat V_M^{ind}:{\cal W} \to \mathbb{R} $ denote the value function of the RMAB under the gain index policy. That is, for any $Y\in \cal W$,
	\begin{align} \label{eq:88}
	\hat V_M^{ind}(Y) = {E_{ind}}\left[ {\sum\limits_{t = 1}^\infty  {{\beta ^{t - 1}}\sum\limits_{i = 1}^M {H\left( {{X_i}(t)} \right)} |Y} } \right] = \sum\limits_{i = 1}^M {{E_{ind}}\left[ {\sum\limits_{t = 1}^\infty  {{\beta ^{t - 1}}H\left( {{X_i}(t)} \right)|Y} } \right]}  \buildrel \Delta \over = \sum\limits_{i = 1}^M {V_i^{ind}(Y)} 
	\end{align}
	where ${V_i^{ind}(Y)}$ is the expected total discounted cost generated by the  $i$-th bandit with the initial state of the RMAB being $Y$. Likewise, we use $\hat V_M^{or}:{\cal W} \to \mathbb{R} $ to represent the value function of the RMAB under the OR policy:
	\begin{align*}
	\hat V_M^{or}(Y) &= {E_{{\pi ^*}}}\left[ {\sum\limits_{t = 1}^\infty  {{\beta ^{t - 1}}\sum\limits_{i = 1}^M {H\left( {{X_i}(t)} \right)} |Y} } \right]\\
	&= {E_{{\pi ^*}}}\left[ {\sum\limits_{t = 1}^\infty  {{\beta ^{t - 1}}\sum\limits_{i = 1}^M {H\left( {{X_i}(t)} \right)}  + {\lambda ^*}\sum\limits_{i = 1}^M {{u_i}(t)} |Y} } \right] - \frac{{m{\lambda ^*}}}{{1 - \beta }}.
	\end{align*}
	Note that the second line is the optimal value of problem P1(c) with initial state $Y$. We then have
	\begin{align}
	J_M^{rel} = \sum\limits_{Y \in {\cal W}} {\zeta (Y)\hat V_M^{or}(Y)} ,\quad J_M^{ind} = \sum\limits_{Y \in {\cal W}} {\zeta (Y)\hat V_M^{ind}(Y)}.
	\end{align}
	Denote by $\mathbf{P}^{or}$ and $\mathbf{P}^{ind}$ the transition matrices of the RMAB under the OR policy and the gain index policy, respectively. In this proof, we use $Y_i$ to denote the state of the $i$-th bandit when the RMAB is in state $Y\in \cal W$, i.e., $Y=[Y_1,\cdots,Y_M]$. In addition, $Y_{1:k}$ denotes the state vector of the first $k$ bandits. The value function will be written as $\hat V_M^{ind}({Y_{1:M - 1}},{Y_M})$ when we need to emphasize the last element of $Y$. According to the Bellman equation,
	\begin{align} \label{eq:Vind-comp}
	\hat V_M^{ind}(Y) = \sum\limits_{i = 1}^M {H({Y_i})}  + \beta \sum\limits_{Y' \in {\cal W}} {{{\bf{P}}^{ind}}(Y'|Y)\hat V_M^{ind}(Y'),\quad Y \in {\cal W}.} 
	\end{align}
	Let $\hat{H}(Y) = \sum_{i=1}^{M}H(Y_i)$. Then, just like in the proof of Theorem \ref{thm:errorbound1}, \eqref{eq:Vind-comp} can be written in vector form as
	\begin{align} \label{eq:Vind-vec}
	\hat V_M^{ind} = \hat H + \beta {{\bf{P}}^{ind}}\hat V_M^{ind}.
	\end{align}
	Similarly, for the OR policy, we have
	\begin{align*}
	\hat V_M^{or} = \hat H + \beta {{\bf{P}}^{or}}\hat V_M^{or}.
	\end{align*}
	Using a similar argument as in the proof of Theorem 2 yields
	\begin{align} \label{eq:Vdiff}
	\hat V_M^{ind} - \hat V_M^{or} = \beta {\left( {{\bf{I}} - \beta {{\bf{P}}^{or}}} \right)^{ - 1}}\left[ {{{\bf{P}}^{ind}}\hat V_M^{ind} - {{\bf{P}}^{or}}\hat V_M^{ind}} \right].
	\end{align}
	Let ${b^{ind}} = {{\bf{P}}^{ind}}\hat V_M^{ind}$. Comparing \eqref{eq:Vind-vec} with \eqref{eq:Vind-comp}, $b^{ind}$ can be expressed in component form as
	\begin{align*}
	{b^{ind}}(Y) = \sum\limits_{Y' \in {\cal W}} {{{\bf{P}}^{ind}}(Y'|Y)\hat V_M^{ind}(Y')} ,\quad Y \in {\cal W}.
	\end{align*}
	Likewise, define ${b^{or}} = {{\bf{P}}^{or}}\hat V_M^{or}$. Consider a state $Y\in \cal W$ in which the actions of the OR policy and the gain index policy differ in only one bandit (without loss of generality, assume they differ in bandit $M$). In particular, let $u^{or}_k$ and $u^{ind}_k$ denote the actions for the  $k$-th bandit taken by the OR policy and the gain index policy in state $Y$, respectively. Then $u^{or}_k =u^{ind}_k$ for $1\le k\le M-1$ and $u^{or}_M \neq u^{ind}_M$. Given the action of each bandit, the one-step state transitions of all bandits are mutually independent. We thus have
	\begin{align*}
	{{\bf{P}}^{or}}(Y'|Y) = \prod\limits_{k = 1}^M {\Pr ({Y'_k}|{Y_k},u_k^{or})} ,\quad {{\bf{P}}^{ind}}(Y'|Y) = \prod\limits_{k = 1}^M {\Pr ({Y'_k}|{Y_k},u_k^{ind}),} \quad Y,Y' \in {\cal W}.
	\end{align*}
	Then
	\begin{align*}
	&{b^{or}}(Y) - {b^{ind}}(Y) = \sum\limits_{Y' \in {\cal W}} {\prod\limits_{k = 1}^{M - 1} {\Pr ({Y'_k}|{Y_k},u_k^{or})\left[ {\Pr ({Y'_M}|{Y_M},u_M^{or}) - \Pr ({Y'_M}|{Y_M},u_M^{ind})} \right]\hat V_M^{ind}(Y')} } \\
	&= \sum\limits_{{Y'_{1:M - 1}}} {\prod\limits_{k = 1}^{M - 1} {\Pr ({Y'_k}|{Y_k},u_k^{or})} \sum\limits_{{Y'_M}} {\left[ {\Pr ({Y'_M}|{Y_M},u_M^{or}) - \Pr ({Y'_M}|{Y_M},u_M^{ind})} \right]\hat V_M^{ind}({Y'_{1:M - 1}},{Y'_M})} }. 
	\end{align*}
	There are two cases: (i) $u_M^{or} = 0,u_M^{ind} = 1$; and (ii) $u_M^{or} = 1,u_M^{ind} = 0$. In both cases, substituting the transition probability of bandit $M$  given by \eqref{eq:SbTrans} into the above equation yields
	\begin{align} \notag
	&\left| {{b^{or}}(Y) - {b^{ind}}(Y)} \right|\\ \notag
	 =& \left| {\sum\limits_{{Y'_{1:M - 1}}} {\prod\limits_{k = 1}^{M - 1} {\Pr ({Y'_k}|{Y_k},u_k^{or})} \rho \left[ {\hat V_M^{ind}({Y'_{1:M - 1}},\tau ) - \sum\limits_{i \in [N]} {{Y_M}(i)\hat V_M^{ind}({Y'_{1:M - 1}},{\bf{T}}_i^{(M)})} } \right]} } \right|\\ \notag
	\le& \rho \mathop {\max }\limits_{{Y'_{1:M - 1}}} \left| {\hat V_M^{ind}({Y'_{1:M - 1}},\tau ) - \sum\limits_{i \in [N]} {{Y_M}(i)\hat V_M^{ind}({Y'_{1:M - 1}},{\bf{T}}_i^{(M)})} } \right|\\ \label{eq:93}
	=& \rho \mathop {\max }\limits_{{Y'_{1:M - 1}}} \left| {\sum\limits_{k = 1}^M {\left[ {V_k^{ind}({{Y'}_{1:M - 1}},\tau ) - \sum\limits_{i \in [N]} {{Y_M}(i)V_k^{ind}({Y'_{1:M - 1}},{\bf{T}}_i^{(M)})} } \right]} } \right|
	\end{align}
	where $\tau  = {{\bf{T}}^{(M)}}{Y_M}$ is the resulting state of bandit $M$ upon taking passive action in $Y_M$, and $Y_M(i)$ is the $i$-th element of $Y_M$ (recall that $Y_M$ is a belief state). The last equality follows from \eqref{eq:88}. As $M\to \infty$, bandit $i$ and bandit $j$ are independent for any $i\neq j$. This is so because any two bandits are weakly coupled (under the gain index policy) via the constraint that exactly $m$ bandits can be selected. As $m$ tends to $\infty$, the evolution of bandit $i$ has no effect on the actions and evolution of bandit $j$. Therefore, for any $k\neq M$,
	\begin{align*}
	V_k^{ind}({Y'_{1:M - 1}},\tau ) = V_k^{ind}({Y'_{1:M - 1}},{\bf{T}}_i^{(M)}),\quad \forall i, \forall{Y'_{1:M - 1}}.
	\end{align*}
	Hence \eqref{eq:93} becomes
	\begin{align*}
	\left| {{b^{or}}(Y) - {b^{ind}}(Y)} \right| \le \rho \mathop {\max }\limits_{{Y'_{1:M - 1}}} \left| {V_M^{ind}({Y'_{1:M - 1}},\tau ) - \sum\limits_{i \in [N]} {{Y_M}(i)V_M^{ind}({Y'_{1:M - 1}},{\bf{T}}_i^{(M)})} } \right|.
	\end{align*}
	Let
	\begin{align*}
	\Delta  \triangleq \rho \mathop {\max }\limits_{{Y_M},{Y'_M}} \mathop {\max }\limits_{{Y'_{1:M - 1}}} \left| {V_M^{ind}({Y'_{1:M - 1}},{Y'_M}) - V_M^{ind}({Y'_{1:M - 1}},{Y_M})} \right|.
	\end{align*}
	Since $V^{ind}_i$ is bounded for every $i\in [M]$, $\Delta$ is also bounded. According to the triangle inequality,  for any ${Y_{1:M}},{Y'_{M - 1}}$ and $Y'_M$,
	\begin{align*}
	&\left| {\hat V_M^{ind}({Y_{1:M }}) - \hat V_M^{ind}({Y_{1:M - 2}},{Y'_{M - 1}},{Y'_M})} \right|\\
	\le& \left| {\hat V_M^{ind}({Y_{1:M }}) - \hat V_M^{ind}({Y_{1:M - 1}},{Y'_{M }})} \right| 
	+ \left| {\hat V_M^{ind}({Y_{1:M - 1}},{Y'_{M }}) - \hat V_M^{ind}({Y_{1:M - 2}},{Y'_{M - 1}},{Y'_M})} \right|\\
	\le& 2\Delta .
	\end{align*}
	In general, if $Y$ is a state in which the actions of the OR policy and the gain index policy differ in $j$  bandits. Then using \eqref{eq:93} and the above argument yields
	\begin{align*}
	\left| {{b^{ind}}(Y) - {b^{or}}(Y)} \right| \le j\Delta.
	\end{align*}
	
	For any positive integer $k$, let $\mathcal{W}_k \subseteq \cal W$ denote the set of states satisfying the following: at any time $t$, if the state of the RMAB is in set $\mathcal{W}_k$, then
	\begin{align*}
	\left| {y_t^{(M)} - m} \right| \le \frac{1}{2}k\sqrt M.
	\end{align*}
	As we discussed around \eqref{eq:dit}, if $\left| {y_t^{(M)} - m} \right| = j$, then the OR policy and the gain index policy take different actions for exactly $j$ bandits. Hence $\mathcal{W}_k$ is the set of states in which the two policies take different actions for at most $k\sqrt{M}/2$ bandits. Therefore, 
	\begin{align} \label{eq:94}
	\left| {{b^{ind}}(Y) - {b^{or}}(Y)} \right| \le \frac{1}{2}k\sqrt M \Delta ,\quad \forall Y \in {{\cal W}_k}.
	\end{align}
	Note that
	\begin{align} \label{eq:95}
	{\left( {{\bf{I}} - \beta {{\bf{P}}^{or}}} \right)^{ - 1}} = \sum\limits_{t = 0}^\infty  {{\beta ^t}{{\left( {{{\bf{P}}^{or}}} \right)}^t}}.
	\end{align}
	Denote by $Y(t)$ the state of the RMAB at time $t$. It follows from \eqref{eq:Vdiff}, \eqref{eq:94}, and \eqref{eq:95} that
	\begin{align} \notag
	&J_M^{ind} - J_M^{rel} =\sum\limits_{t = 1}^\infty  {{\beta ^t}\sum\limits_{Y \in {\cal W}} {\left[ {{b^{ind}}(Y) - {b^{or}}(Y)} \right]{P^{or}}\left\{ {Y(t) = Y|Y(1) \sim \zeta } \right\}} } \\ \notag
	=& \sum\limits_{t = 1}^\infty  {{\beta ^t}\sum\limits_{Y \in {\cal W}} {\left[ {{b^{ind}}(Y) - {b^{or}}(Y)} \right]\zeta (Y)} } \\ \label{eq:96}
	=& \sum\limits_{t = 1}^\infty  {{\beta ^t}\sum\limits_{Y \in {{\cal W}_k}} {\left[ {{b^{ind}}(Y) - {b^{or}}(Y)} \right]\zeta (Y)} }  + \sum\limits_{t = 1}^\infty  {{\beta ^t}\sum\limits_{Y \in {\cal W} - {{\cal W}_k}} {\left[ {{b^{ind}}(Y) - {b^{or}}(Y)} \right]\zeta (Y)} } 
	\end{align}
	where $P^{or}\{\cdot | \cdot \}$ represents the state distribution of the RMAB under the OR policy. The second line follows from the fact that $\zeta$ is the equilibrium distribution of the RMAB governed by the OR policy. According to Proposition 1,
	\begin{align}
	\sum\limits_{Y \in {{\cal W}_k}} {\zeta (Y)}  \ge \Phi (k) - \Phi ( - k).
	\end{align}
	For large $k$, we can approximate $1-\Phi (k) + \Phi ( - k)$ by
	\begin{align*}
	\sum\limits_{Y \in {\cal W} - {{\cal W}_k}} {\zeta (Y)}  \le 1 - \Phi (k) + \Phi ( - k) \approx \frac{{{e^{ - {k^2}/2}}}}{{k\sqrt {\pi /2} }}.
	\end{align*}
	Let $k=M^{1/4}$, we can verify that the second term of \eqref{eq:96} tends to 0 as $M\to \infty$. We thus write it as $O(M)$ for simplicity. It then follows from \eqref{eq:94} and \eqref{eq:96} that
	\begin{align*}
	\frac{1}{M}\left| {J_M^{ind} - J_M^{rel}} \right| &\le \frac{1}{{2M}}\sum\limits_{t = 1}^\infty  {{\beta ^t}\sum\limits_{Y \in {{\cal W}_k}} {{M^{\frac{3}{4}}}\Delta \zeta (Y)} }  + \frac{1}{M}O(M)\\
	&\le \frac{1}{{2{M^{1/4}}}}\frac{{\beta \Delta }}{{1 - \beta }} + \frac{1}{M}O(M).
	\end{align*}
	Since $\Delta$ is bounded for all $M$, we have
	\begin{align*}
	\mathop {\lim }\limits_{M \to \infty } \frac{1}{M}\left| {J_M^{ind} - J_M^{rel}} \right| = 0{\rm{ }}.
	\end{align*}
	The desired result is obtained using the fact that $J_M^{ind} \ge J_M^{opt} \ge J_M^{rel}$.	
	
\end{IEEEproof}

\subsection*{D. Proof of Theorem \ref{thm:ave-opt}}
This part provides a short proof for the asymptotic optimality of the gain index policy for the average cost problem. Since the proof is similar to that for the discounted cost problem (Theorem \ref{thm:optimality}), we restrict our attention to pointing out and addressing the technique issues that are different from the discounted cost problem. Repetitive steps will be omitted. 
\begin{IEEEproof}
	Let $\hat Z_M^{ind}:{\cal W} \to \mathbb{R} $ denote the differential value function of the RMAB under the gain index policy. Then the relationship between $\hat Z_M^{ind}$ and $G_M^{ind}$ can be expressed in vector form as
	\begin{align} \label{eq:Zhat}
	\hat Z_M^{ind} + G_M^{ind}e = \hat{H} + {{\bf{P}}^{ind}}\hat Z_M^{ind},
	\end{align}
	where $e$ denotes the all-one vector. It is well known that $\hat Z_M^{ind}(Y)$  can be expressed as
	\begin{align} \notag
	\hat Z_M^{ind}(Y) &= {E_{ind}}\left[ {\sum\limits_{t = 0}^\infty  {\left( {\sum\limits_{i = 1}^M {H({X_i}(t)) - G_M^{ind}} } \right)|Y} } \right]   \\ \label{eq:99}
	&= \sum\limits_{i = 1}^M {{E_{ind}}\left[ {\sum\limits_{t = 0}^\infty  {\left( {H({X_i}(t)) - g_i^{ind}} \right)|Y} } \right]}  \buildrel \Delta \over = \sum\limits_{i = 1}^M {Z_i^{ind}(Y)} 
	\end{align}
	where $g^{ind}_i$ is the long-term average cost of the  $i$-th bandit under the gain index policy. The second equality holds because $G_M^{ind} = \sum\nolimits_{i = 1}^M {g_i^{ind}} $. Note that $Z_i^{ind}(Y)$ is bounded for all $Y$ and $i$. Given the equilibrium distribution $\zeta$ of the OR policy, we have $G^{rel}_M = \zeta^\top \hat{H}$ and ${\zeta ^ \top } = {\zeta ^ \top }{{\bf{P}}^{or}}$. Multiplying both sides of \eqref{eq:Zhat} on the left by $\zeta^\top$ yields
	\begin{align} \label{eq:100}
	G_M^{ind} - G_M^{rel} = {\zeta ^ \top }\left( {{{\bf{P}}^{ind}} - {{\bf{P}}^{or}}} \right)\hat Z_M^{ind}.
	\end{align}
	Let ${\bar b^{ind}} = {{\bf{P}}^{ind}}\hat Z_M^{ind}$ and ${\bar b^{or}} = {{\bf{P}}^{or}}\hat Z_M^{ind}$. Using \eqref{eq:99} and the boundedness of $Z_i^{ind}(Y)$, we can prove via the same procedure as in the proof of Theorem \ref{thm:optimality} that
	\begin{align*}
	\mathop {\lim }\limits_{M \to \infty } \frac{1}{M}\left| {G_M^{ind} - G_M^{rel}} \right| = 0.
	\end{align*}
	The desired result follows immediately. We omit the details to reduce mechanical repetitions. Note that $\zeta$ shows up in \eqref{eq:100} without introducing Assumption 1. By contrast, $J_M^{ind} - J_M^{rel}$ relies on Assumption 1 to achieve a similar form (see \eqref{eq:96}). This difference between the two settings allows us to remove Assumption 1 in the average cost problem.
	
\end{IEEEproof}

\bibliographystyle{IEEEtran}
\bibliography{reference}

% Generated by IEEEtran.bst, version: 1.14 (2015/08/26)
\begin{thebibliography}{10}
\providecommand{\url}[1]{#1}
\csname url@samestyle\endcsname
\providecommand{\newblock}{\relax}
\providecommand{\bibinfo}[2]{#2}
\providecommand{\BIBentrySTDinterwordspacing}{\spaceskip=0pt\relax}
\providecommand{\BIBentryALTinterwordstretchfactor}{4}
\providecommand{\BIBentryALTinterwordspacing}{\spaceskip=\fontdimen2\font plus
\BIBentryALTinterwordstretchfactor\fontdimen3\font minus
  \fontdimen4\font\relax}
\providecommand{\BIBforeignlanguage}[2]{{%
\expandafter\ifx\csname l@#1\endcsname\relax
\typeout{** WARNING: IEEEtran.bst: No hyphenation pattern has been}%
\typeout{** loaded for the language `#1'. Using the pattern for}%
\typeout{** the default language instead.}%
\else
\language=\csname l@#1\endcsname
\fi
#2}}
\providecommand{\BIBdecl}{\relax}
\BIBdecl

\bibitem{AoI_2011}
S.~{Kaul}, M.~{Gruteser}, V.~{Rai}, and J.~{Kenney}, ``Minimizing age of
  information in vehicular networks,'' in \emph{8th Annual IEEE Communications
  Society Conference on Sensor, Mesh and Ad Hoc Communications and Networks},
  2011, pp. 350--358.

\bibitem{AoI_originalpaper}
S.~{Kaul}, R.~{Yates}, and M.~{Gruteser}, ``Real-time status: How often should
  one update?'' in \emph{Proceedings IEEE INFOCOM}, 2012, pp. 2731--2735.

\bibitem{AoIdesign_Kadota2019}
I.~{Kadota} and E.~{Modiano}, ``Minimizing the age of information in wireless
  networks with stochastic arrivals,'' \emph{IEEE Transactions on Mobile
  Computing}, vol.~20, no.~3, pp. 1173--1185, 2021.

\bibitem{AoI2016Costa}
M.~{Costa}, M.~{Codreanu}, and A.~{Ephremides}, ``On the age of information in
  status update systems with packet management,'' \emph{IEEE Transactions on
  Information Theory}, vol.~62, no.~4, pp. 1897--1910, 2016.

\bibitem{AoICoding_2022TII}
E.~Najm, E.~Telatar, and R.~Nasser, ``Optimal age over erasure channels,''
  \emph{IEEE Transactions on Information Theory}, vol.~68, no.~9, pp.
  5901--5922, 2022.

\bibitem{AoIdesign_henry2020}
J.~{Li}, Y.~{Zhou}, and H.~{Chen}, ``Age of information for multicast
  transmission with fixed and random deadlines in {IoT} systems,'' \emph{IEEE
  Internet of Things Journal}, vol.~7, no.~9, pp. 8178--8191, 2020.

\bibitem{sombabu2022whittle}
B.~Sombabu, B.~Dedhia, and S.~Moharir, ``Whittle index based age-of-information
  aware scheduling for markovian channels,'' \emph{Computer Networks and
  Communications}, vol.~1, no.~1, pp. 59--84, 2022.

\bibitem{AoI_Cao2023}
X.~Cao, J.~Wang, Y.~Cheng, and J.~Jin, ``Optimal sleep scheduling for
  energy-efficient {AoI} optimization in industrial internet of things,''
  \emph{IEEE Internet of Things Journal}, pp. 1--1, 2023.

\bibitem{AoIcost_nonlinear_MIT}
V.~Tripathi and E.~Modiano, ``A whittle index approach to minimizing functions
  of age of information,'' in \emph{57th Annual Allerton Conference on
  Communication, Control, and Computing (Allerton)}.\hskip 1em plus 0.5em minus
  0.4em\relax IEEE Press, 2019, pp. 1160--1167.

\bibitem{AoIcost_nonlinear2019}
M.~{Klügel}, M.~H. {Mamduhi}, S.~{Hirche}, and W.~{Kellerer}, ``{AoI}-penalty
  minimization for networked control systems with packet loss,'' in \emph{IEEE
  INFOCOM 2019 - IEEE Conference on Computer Communications Workshops (INFOCOM
  WKSHPS)}, 2019, pp. 189--196.

\bibitem{Aoicost_nonlinear}
J.~P. Champati, M.~H. Mamduhi, K.~H. Johansson, and J.~Gross, ``Performance
  characterization using {AoI} in a single-loop networked control system,''
  \emph{CoRR}, vol. abs/1901.06694, 2019.

\bibitem{AoIaware18}
P.~R. {Jhunjhunwala} and S.~{Moharir}, ``Age-of-information aware scheduling,''
  in \emph{International Conference on Signal Processing and Communications
  (SPCOM)}, 2018, pp. 222--226.

\bibitem{UoI_Gongpu}
G.~Chen, S.~C. Liew, and Y.~Shao, ``Uncertainty-of-information scheduling: A
  restless multiarmed bandit framework,'' \emph{IEEE Transactions on
  Information Theory}, vol.~68, no.~9, pp. 6151--6173, 2022.

\bibitem{gittins_RMAB_book}
J.~Gittins, K.~Glazebrook, and R.~Weber, \emph{Multi-armed bandit allocation
  indices, 2nd Edition}.\hskip 1em plus 0.5em minus 0.4em\relax John Wiley \&
  Sons, 2011.

\bibitem{MutualInform_sunyin}
Y.~{Sun} and B.~{Cyr}, ``Information aging through queues: {A} mutual
  information perspective,'' in \emph{IEEE 19th International Workshop on
  Signal Processing Advances in Wireless Communications (SPAWC)}, 2018, pp.
  1--5.

\bibitem{VoI_HiddenMC}
Z.~Wang, M.-A. Badiu, and J.~P. Coon, ``A framework for characterizing the
  value of information in hidden markov models,'' \emph{IEEE Transactions on
  Information Theory}, vol.~68, no.~8, pp. 5203--5216, 2022.

\bibitem{VoI_entropy}
T.~Soleymani, S.~Hirche, and J.~S. Baras, ``Optimal self-driven sampling for
  estimation based on value of information,'' in \emph{2016 13th International
  Workshop on Discrete Event Systems (WODES)}, 2016, pp. 183--188.

\bibitem{Leiyin2020}
W.~Wang and L.~Ying, ``Learning parallel markov chains over unreliable wireless
  channels,'' in \emph{2020 54th Annual Conference on Information Sciences and
  Systems (CISS)}, 2020, pp. 1--6.

\bibitem{Coding_MarkovSource}
S.~Poojary, S.~Bhambay, and P.~Parag, ``Real-time status updates for markov
  source,'' \emph{IEEE Transactions on Information Theory}, vol.~65, no.~9, pp.
  5737--5749, 2019.

\bibitem{Sampling_Markov}
J.~P. Champati, M.~Skoglund, M.~Jansson, and J.~Gross, ``Detecting state
  transitions of a markov source: Sampling frequency and age trade-off,''
  \emph{IEEE Transactions on Communications}, vol.~70, no.~5, pp. 3081--3095,
  2022.

\bibitem{AoII_maatouk2020}
A.~Maatouk, S.~Kriouile, M.~Assaad, and A.~Ephremides, ``The age of incorrect
  information: A new performance metric for status updates,'' \emph{IEEE/ACM
  Transactions on Networking}, vol.~28, no.~5, pp. 2215--2228, 2020.

\bibitem{RMAB_PSPACEhard}
C.~H. Papadimitriou and J.~N. Tsitsiklis, ``The complexity of optimal queuing
  network control,'' \emph{Mathematics of Operations Research}, vol.~24, no.~2,
  pp. 293--305, 1999.

\bibitem{RMAB_Whittle1988}
P.~Whittle, ``Restless bandits: Activity allocation in a changing world,''
  \emph{Journal of Applied Probability}, vol.~25, pp. 287--298, 1988.

\bibitem{weber1990index}
R.~R. Weber and G.~Weiss, ``On an index policy for restless bandits,''
  \emph{Journal of applied probability}, vol.~27, no.~3, pp. 637--648, 1990.

\bibitem{nino_2001}
J.~Ni{\~n}o-Mora, ``Restless bandits, partial conservation laws and
  indexability,'' \emph{Advances in Applied Probability}, vol.~33, no.~1, pp.
  76--98, 2001.

\bibitem{nino2007dynamic}
------, ``Dynamic priority allocation via restless bandit marginal productivity
  indices,'' \emph{Top}, vol.~15, no.~2, pp. 161--198, 2007.

\bibitem{Whittle_app2006}
K.~D. Glazebrook, D.~Ruiz-Hernandez, and C.~Kirkbride, ``Some indexable
  families of restless bandit problems,'' \emph{Advances in Applied
  Probability}, vol.~38, no.~3, pp. 643--672, 2006.

\bibitem{Shiling2020}
J.~{Wang}, X.~{Ren}, Y.~{Mo}, and L.~{Shi}, ``Whittle index policy for dynamic
  multichannel allocation in remote state estimation,'' \emph{IEEE Transactions
  on Automatic Control}, vol.~65, no.~2, pp. 591--603, 2020.

\bibitem{Liu2010}
K.~{Liu} and Q.~{Zhao}, ``Indexability of restless bandit problems and
  optimality of whittle index for dynamic multichannel access,'' \emph{IEEE
  Transactions on Information Theory}, vol.~56, no.~11, pp. 5547--5567, 2010.

\bibitem{Whittle_app2018}
K.~E. {Avrachenkov} and V.~S. {Borkar}, ``Whittle index policy for crawling
  ephemeral content,'' \emph{IEEE Transactions on Control of Network Systems},
  vol.~5, no.~1, pp. 446--455, 2018.

\bibitem{villar_2016}
S.~S. Villar, ``Indexability and optimal index policies for a class of
  reinitialising restless bandits,'' \emph{Probability in the Engineering and
  Informational Sciences}, vol.~30, no.~1, pp. 1--23, 2016.

\bibitem{Index_compute2020}
N.~Akbarzadeh and A.~Mahajan, ``Restless bandits: indexability and computation
  of whittle index,'' \emph{arXiv preprint arXiv:2008.06111}, 2020.

\bibitem{puterman1994markov}
M.~L. Puterman, \emph{Markov decision processes: discrete stochastic dynamic
  programming}.\hskip 1em plus 0.5em minus 0.4em\relax John Wiley \& Sons,
  1994.

\bibitem{bertsekas1997nonlinear}
D.~P. Bertsekas, ``Nonlinear programming,'' \emph{Journal of the Operational
  Research Society}, vol.~48, no.~3, pp. 334--334, 1997.

\bibitem{boyd2004convex}
S.~Boyd and L.~Vandenberghe, \emph{Convex optimization}.\hskip 1em plus 0.5em
  minus 0.4em\relax Cambridge university press, 2004.

\bibitem{Kadota2018TON}
I.~Kadota, A.~Sinha, E.~Uysal-Biyikoglu, R.~Singh, and E.~Modiano, ``Scheduling
  policies for minimizing age of information in broadcast wireless networks,''
  \emph{IEEE/ACM Transactions on Networking}, vol.~26, no.~6, pp. 2637--2650,
  2018.

\bibitem{dutta1991discounted}
P.~K. Dutta, ``What do discounted optima converge to?: A theory of discount
  rate asymptotics in economic models,'' \emph{Journal of Economic Theory},
  vol.~55, no.~1, pp. 64--94, 1991.

\end{thebibliography}

\end{document}